\documentclass[oneside,12pt]{article}
\usepackage{amsfonts,amsmath,amssymb,mathrsfs}
\usepackage{dsfont}
\usepackage{graphicx}
\usepackage{bm}
\begin{document}
\title{On Adiabatic Pair Creation}

\author{Peter Pickl\footnote{Institut f\"ur theoretische Physik, Universit\"{a}t
        Wien, Boltzmanngasse 5, 1090 Vienna, Austria
         E-mail: pickl@mathematik.uni-muenchen.de}, Detlef Duerr\footnote{Mathematisches Institut der Universit\"{a}t
         M\"{u}nchen, Theresienstra{\ss}e 39, 80333 M\"{u}nchen, Germany.
         E-mail: duerr@mathematik.uni-muenchen.de}}
\date{\today}




\newtheorem{thm}{Theorem}[section]
\newtheorem{cor}[thm]{Corollary}
\newtheorem{lem}[thm]{Lemma}
\newtheorem{proof}[thm]{Proof}
\newtheorem{prop}[thm]{Proposition}
\newtheorem{defn}[thm]{Definition}
\newtheorem{rem}[thm]{Remark}
\newtheorem{con}[thm]{Condition}
\newtheorem{nota}[thm]{Notation}

\newcommand{\ed}{\bf}
\newcommand{\eed}{\rm}

\sloppy

\maketitle
\begin{abstract}
We give here the proof that pair creation arises from the Dirac
equation with an external time dependent potential. Pair creation
happens with probability one if the potential changes
adiabatically in time and becomes overcritical, that is when an
eigenvalue curve (as function of time) bridges the gap between the
negative and positive spectral continuum. The potential may be
assumed to be zero at large negative and large positive times. The
rigorous treatment of this effect has been lacking since the
pioneering work of Beck, Steinwedel and S\"u\ss mann \cite{beck}
in 1963 and Gershtein and Zeldovich \cite{gersh} in 1970.
\end{abstract}

\newpage

\tableofcontents

\newpage

\section{Introduction}

Adiabatic pair creation (APC)  has been called---unfortunately
misleading---spontaneous pair creation
(\cite{greiner},\cite{mueller} -
\cite{diss},\cite{prodan,PRLgreiner2,reinhardt,scharf,exp1,PRLgreiner,thaller}).
The creation of  electron positron pairs in very strong external
classical electromagnetic fields arises straight forwardly from
the Dirac sea interpretation of negative energy states. After
Dirac\cite{diracbook} it has been discussed as an academic problem
by Klein \cite{klein}, Sauter \cite{sauter} , Heisenberg and Euler
\cite{heis}, Schwinger \cite{schwinger} and Brezin and Itzykson
\cite{itzyk}. A more realistic setting was hinted at by Beck,
Steinwedel and S\"u\ss mann, \cite{beck} and worked out by
Gershtein and Zeldovich \cite{gersh} as APC. In the common physics
language it may be described as follows:
 An adiabatically increasing  electric potential lifts a particle
from the sea to the positive energy subspace where it scatters and
when the potential is gently switched off one has one free electron
and one unoccupied state---a hole---in the sea. The experimental
verification needs very strong classical fields \cite{greiner} and
is discussed elsewhere \cite{PDPhysrev}. In this respect we would
like to remark
 that a coherent analysis of the existence of APC has
been lacking until recently \cite{diss}.   Earlier quantitative
results based on an ad hoc and incoherent analysis (see for
example \cite{mprg}, \cite{PRLgreiner}) are false concerning the
rate and the outgoing momenta of the spontaneously created pairs
(see \cite{PDPhysrev}). There have been also results in the
mathematical physics literature related to APC, notably
\cite{nenciu1, nenciu2, prodan} but those results do not come to
grasp at all with the heart of the problem of APC, which is the
control of the wavefunction evolution within the neighborhood of
the spectral edge $mc^2$.

In APC one considers  the so called external field problem, where
interactions between the charges are neglected. Vacuum polarization
will in general perturb the external field
 and - using mean field approximation -(see \cite{hainzl}) one may think of
the external field as an effective field.

The existence of APC in second quantized external field Dirac
theory (if the latter exists\footnote{It is well known that the
lifting of the Dirac evolution (with a smooth field of compact
support) to Fock space  (second quantisation) is possible if and
only if the Shale-Stinespring condition is satisfied
\cite{thaller, scharf}, which is the case if and only if the
magnetic field vanishes. On the other hand, the S-matrix can
always be lifted.} is equivalent to the existence of certain types
of solutions  of the Dirac equation (see e.g. \cite{nenciu1} and
\cite{diss}) which we describe below. The existence of APC in
terms of the second quantised S-matrix theory of the Dirac
equation with external field  is ``by definition'' equivalent to
the existence of these types of solutions of the Dirac equation.
We shall in fact formulate our result in terms of the solutions of
the Dirac equation and use the Dirac sea picture for the
interpretation of the particular solution we prove in this paper
to exist.

 Consider the Dirac equation
with external electric field. Then the potential $A$ can be chosen
as a real valued multiple of the $4\times 4$ unit matrix. (We wish
to note that the results can be extended to general four
potentials. Concerning strong magnetic fields we wish to call
attention to the recent work of Dolbeault et.al. \cite{loss} as
well as \cite{PDPhysrev}). $A mc^2$ gives the potential in the
units $eV$. We assume that the potential $A$ varies slowly with
time, expressed by $A_{\varepsilon \tau}$, where $\varepsilon$ is
a dimensionless small parameter (given by the physics, see
\cite{PDPhysrev} for some examples) which in this work will
eventually be sent to zero to obtain limit results. Here
$\tau=\frac{mc^2}{\hbar}t$  and
$\mathbf{x}=\frac{mc}{\hbar}\mathbf{r}$ are the dimensionless
microscopic time- and space-scales  and the Dirac equation  in the
standard representation
 reads with the notation
$\mathbf{x}=(x_1,x_2,x_3)\in\mathbb{R}^3$ and $
\partial_{l}:=\frac{\partial}{\partial_{x_l}}$
\begin{eqnarray}\label{Diracmicalt}
   i\frac{\partial\psi_{\tau}(\mathbf{x})}{\partial \tau}&=&-i
   \sum_{l=1}^{3}\alpha_{l}\partial_{l}\psi_{\tau}(\mathbf{x})+A_{\varepsilon \tau}(\mathbf{x})\psi_{\tau}+\beta
   \psi_{\tau}(\mathbf{x})
   \nonumber\\&\equiv&(D_0+A_{\varepsilon
   \tau}(\mathbf{x}))\psi_{\tau}\,.
\end{eqnarray}
We introduce in (\ref{Diracmicalt}) the macroscopic time scale
$s=\varepsilon\tau$. We wish to restrict ourselves to potentials
$A_s$ which can be factorized into a space- and a time dependent
factor $A_s(\mathbf{x}):=A(\mathbf{x})\mu(s)$, a restriction of
technical nature which eases notations and computations and which
furthermore helps to picture a spatial potential well which
changes its depth with time. It is helpful to have this picture in
mind, because the potential does act as an elevator, as we shall
explain below.
 We thus have
\begin{eqnarray}\label{Diracmic}
   i\frac{\partial\psi_{s}}{\partial s}\equiv \frac{1}{\varepsilon}
   (D_0+A\mu(s))\psi_{s}\equiv\frac{1}{\varepsilon}D_{\mu(s)}\psi_{s}\;.
\end{eqnarray}
 Furthermore we wish to restrict ourselves to
potentials $A_s(\mathbf{x})$ which are smooth, bounded, compactly
supported in $\mathbf{x}$ and $s$ and positive.
%
The spectrum of the  free Dirac operator $D_0$ is absolutely
continuous and given by $(-\infty,-1]\cap[1,\infty)$, defining
``negative and positive energy'' subspaces.  In the Dirac sea
interpretation wavefunctions which lie in the positive energy
subspace of the free Dirac operator are interpreted as wavefunctions
of electrons. The so called vacuum of second quantized Dirac
equation corresponds in the Dirac sea picture to all ``states of the
negative energy subspace being occupied by particles''---the Dirac
sea. ``Holes'' in the Dirac sea are unoccupied negative energy
states which are interpreted as anti-electrons, i.e. positrons.
\begin{figure}[t]\label{bild}
\begin{center}
\leavevmode %
\includegraphics[width=.6\textwidth]{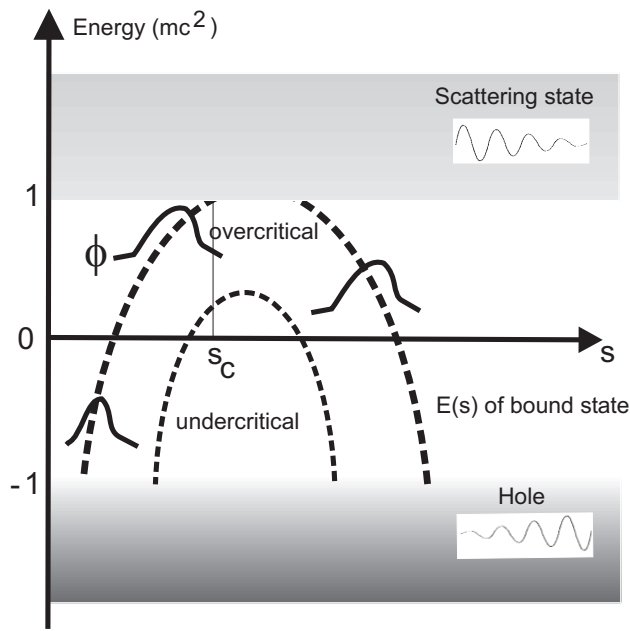}
\caption[Pair]{\footnotesize Schematic presentation of the
adiabatic pair creation. It shows the spectrum of the Dirac
operator $D_{\mu(s)}$ as function of $s$. Depending on the
strength of the potential there may exist bound state energy
curves $E(s)$, one or more of which may bridge the spectral gap
(overcritical case). Also schematically drawn  are bound states
$\Phi$ at various undercritical times. No bound states exist in
the lower and upper spectral continua $(-\infty,-1)$ and $(
1,\infty)$. Pair creation is achieved (with probability one) if a
particle from the sea which occupied at small times $s$ the bound
states $\Phi$ corresponding to the gap bridging bound state energy
curve scatters after the bound state curve has reached the upper
spectral set $[1,\infty)$ at time $s_c$, and when that bound state
becomes
 a scattering state. The ``returning bound state'' is then
unoccupied producing a hole in the sea.}
\end{center}
\end{figure}\nopagebreak
 The goal of our paper is to assert that there exist solutions of
(\ref{Diracmic}) which describe pair creation. We explain what
that means.

The main idea of APC, as illustrated in figure \ref{bild} is as
follows. Consider first the spectrum of the time dependent Dirac
operator $D_{\mu(s)}$. At large negative and large positive times
when $A_s=0$, $D_{\mu(s)}=D_0$ and  we have the spectrum of $D_0$.
At times at which $A_s\ne0$ there may be eigenvalues in the gap
$[-1,1]$, while the continuous spectrum remains unchanged. The
eigenvalues change with the strength of the potential, i.e. with
time $s$ (bound state energy curve $E(s)$ in figure \ref{bild}).
Suppose first that no eigenvalue  reaches $1$, .i.e. no bound
state energy curve bridges the gap (undercritical case).
 The adiabatic theorem (see e.g.
\cite{stefan}) ensures that there is no tunnelling across spectral
gaps meaning that the bound states stay more or less intact when
the potentials changes adiabatically. In terms of solutions of the
Dirac equation (\ref{Diracmic}) that means the following: There
exists no solution when $\varepsilon$ goes to zero which  for
large negative times lies completely within the negative
continuous energy subspace and for large positive times has parts
in the positive continuous energy subspace. In the Dirac sea
interpretation that means: The probability of creating a pair is
zero. No APC.

However, when the external field becomes overcritical (at time
$s_c$ in figure \nolinebreak \ref{bild}),  the highest lying
eigenvalue curve reaches the positive continuum and  the bound
state ceases to exist and becomes a continuum state (a
``resonance'') in the positive continuum subspace. Then there
exists a solution of the Dirac equation which follows
adiabatically the path of this bound state, which for large
negative times must develop into a wavefunction which lies
entirely in the negative continuum energy subspace and for
positive times may have a part in the positive continuum energy
subspace. As indicated in the figure \ref{bild} when the potential
decreases with increasing time there is again a bound state energy
curve bridging the gap. In principle the solution of the Dirac
equation can have a part which ``follows'' the bound state back
into the negative continuous energy subspace and remains there
when the potential is switched off. In the Dirac sea
interpretation such a solution of the Dirac equation would
correspond to pair creation with a probability determined by the
absolute squares of the parts of the wave functions. We show
however, that no such ``back sliding'' is adiabatically possible,
i.e. no such solution of the Dirac equation exists. The former
bound state scatters in the positive continuum energy subspace,
i.e. it stays there for all later times. In the Dirac sea picture
the ``returning'' bound state remains for sure empty and upon
becoming a state within the negative continuum energy subspace
there is now an unoccupied state in the sea: APC is accomplished
with probability one. One pictorial way to describe APC is to
imagine the potential acting as an elevator, lifting a particle
from the sea to the ``upper''(positive) continuum.
  The scenario
is symmetric under change of sign of the potential: It then
transports an unoccupied state (a hole) from the positive continuum
to the sea and catches a particle from the sea when it is switched
off. The hole (positron) then scatters.

We understand now the type of solution of (\ref{Diracmic}) we wish
to study, namely  one which at some
 time at which an overcritical bound state exists equals that
 bound state.  The scenario we described translates mathematically
 into the task to establish  scattering
of such solutions of the time-inhomogenous Dirac equation
(\ref{Diracmic}). To show to what extend the scenario of APC holds
one must control first that the bound states stay on the adiabatic
time scale intact until the eigenvalues reach the positive
continuum. That is content of an adiabatic lemma without a gap and
``relatively easy'' to establish. The solution of (\ref{Diracmic})
is thus adiabatically essentially represented by the ``time
dependent bound states'' until that time.
 Then we must control the propagation of the
wavefunction (the resonance) emerging from the bound state during
over-criticality. We wish to show that it scatters. This task is
on the one hand far from being easy, since the Dirac operator
changes with time. The time evolution will be controlled by
generalized eigenfunctions, i.e. by the stationary phase argument,
which is of course not standard because the generalized
eigenfunctions themselves are depending now on time. But more than
that on the other hand we must take into account the bad
(resonant) behavior of the generalized eigenfunctions near
criticality. (We wish to note that also \cite{rodni} is concerned
with the wavefunction propagation for time dependent Hamiltonians
but under generic smallness assumptions on the potential,
assumptions which are not fulfilled in our problem). They become
unbounded for critical $k$-values (which are small) and hence the
situation is very much different from the usual scattering
situation governed by ``plane waves'' (see \cite{PDPhysrev} for a
heuristic argument giving some intuition). As we shall find out,
the decay time of a wave, say from a bounded spatial region (i.e.
the time in which roughly half of the mass left the region), is
now (on the microscopic time) $t\sim \varepsilon^{-2/3}$ as
compared to $t\sim\mathcal{O}(1) $ in the common plane wave
scattering situation. This means that the resonance lingers around
the range of the potential for a much longer time than in the
usual scattering of wavefunctions. Such a metastable state decay
has already been suggested by \cite{beck}.

We shall give in the next section the result: Theorem
\ref{onecrit} and Corollary \ref{psipropneu}. The rest of the
paper is devoted to the proof of the theorem. The proof is
technically very involved. Instead of describing here what is in
the sections to follow we first give the result and then give in
Section \ref{skeleton} a skeleton of the proof with a description
of the contents of the sections.

\section{The Result}
We begin with
\begin{nota}
The functions we mainly consider are spinors in the space
$L^n(\mathbb{R}^3,\mathbb{C}^4)$, $n=1,2,\infty$. We shall denote
this space if no ambiguity arises simply by $L^n$. We shall have
two scalar products: (i) For $a,b\in\mathbb{C}^4$:
$\overline{a}b:=\sum_{j=1}^4a_j^*b_j$ where $^* $ denotes complex
conjugation. (ii) For $\psi,\chi\in L^2$:
$\langle\psi,\chi\rangle:=\int\overline{\psi}(\mathbf{x})\chi(\mathbf{x})
d^3x$, $\|\psi\|=\sqrt{\langle\psi,\psi\rangle}$. Warning:
Constants appearing in estimates will generically be denoted by
$C$. We shall not distinguish constants appearing in a sequence of
estimates, i.e. in $X\le C Y\le C Z$ the constants may  differ.
\end{nota}

In the following we will only consider potentials which are
 bounded, compactly supported,  positive and purely electric.
 The latter implies that $A$ will be a multiple of the unit matrix
 (since we stick to one inertial frame throughout the paper). Thus $A$ can be written as a scalar function.  To have
the possibility of pair creation the external (scalar) field
$\mu(s)A$ has to become critical for some time $s$ and the first
such time will be set $s=0$ and we choose $\mu(0)=1$, i.e. $A$ is
critical. Criticality means for us that $D_0+A$ has only bound
states solutions (i.e. $L^2$-solutions and no resonances with energy
$1$) of
\begin{equation}\label{ewg}
(D_0+A)\Phi=\Phi\,.
\end{equation}
 This is the generic case
(see e.g. \cite {klaus}) of critical potentials in the Dirac
equation.  We shall now collect the conditions in a form most
convenient for our considerations. 

\begin{con}\label{cond} For $A:\mathbb{R}^3\to\mathbb{R}^+$ and $\mu:\mathbb{R}\to \mathbb{R} $  we shall
require that \begin{itemize} \item[(i)]
$A$ has compact support $\mathcal{S}_A$; $A,\nabla A$ are bounded
and $A$ is critical. Furthermore $D_1 = D_0 +A$ has no resonances
for $E=1$ and $E=1$ is n-fold degenerate for some $n\in\mathbb{ N}$
with eigenspace denoted  by $\mathcal{N}$:
\begin{equation}\label{defmengen}
\mathcal{N}:=\{\Phi\in L^2:(D_0+A-1)\Phi=0\}\;.
\end{equation}

 \item[(ii)]For
 any $\mu\in[0,1]$ there exist not more than one eigenvalue
$E_\mu$ of the operator $D_\mu=D_0+\mu A$. Warning: We shall use
the symbol $\mu$ as fixed parameter and as function $\mu(s)$.
 \item[(iii)]
$\mu:\mathbb{R}\to \mathbb{R} $ is continuously differentiable,
its derivative  $\mu'$ is bounded, $\mu(0)=1$ and $\mu'(0)>0$.
  There exists $s_i<0$ and $s_f>0$ such that $\mu(s)=0$ if $s<s_i$ or $s>s_f$.
\end{itemize}
\end{con}
\begin{rem} The condition above is fulfilled by a large class of critical
potentials $A$. (i) is fulfilled for the ground state and generically for excited states (see \cite{klaus}).

(ii) excludes the possibility of having more than one bound state curve entering the upper spectral edge. This
assumption is not essential but makes the proof less heavy.

\end{rem}

Under this condition (see e.g \cite{thaller}) the operator of
interest defining (\ref{Diracmic}) namely
$\frac{1}{\varepsilon}D_{\mu(s)}$ generates a unitary time
evolution denoted by $U^\varepsilon(s,s_0)$ given by
\begin{equation}\label{U}
i\partial_s U^\varepsilon(s,s_0)= \frac{1}{\varepsilon}D_{\mu(s)}
U^\varepsilon(s,s_0)\,,
\end{equation}
generating solutions of (\ref{Diracmic}). The following theorem
and its corollary  assert that there exists a scattering solution
of the Dirac equation (\ref{Diracmic}) which at large negative
times is element of the negative energy spectral subspace of the
free Dirac operator ($A=0$) and at large positive times it is an
element of the positive energy spectral subspace of the free Dirac
operator.
\begin{thm}\label{onecrit}(Decay of the Bound States)\\
Assume condition \ref{cond}. Let $\Phi_{\mu(s_0)}$ be a bound
state of $D_{\mu(s_0)}$ for some $s_0 \in(s_i,0]$. Let
$U^\varepsilon(s,s_0)$ be given by (\ref{U}), i.e.
$\psi_{s}^\varepsilon=U^\varepsilon(s,s_0)\Phi_{\mu(s_0)}$ is the
solution of the Dirac equation (\ref{Diracmic}) with
$\psi_{s_0}^\varepsilon=\Phi_{\mu(s_0)}$. Then for all $\chi\in
L^2$

\begin{equation}\label{onecriteq}
\lim_{\varepsilon\to0}\langle\psi_{s}^\varepsilon,\chi\rangle=0
\end{equation}
for any $s>0$.

\end{thm}

As a Corollary of Theorem \ref{onecrit} and the adiabatic theorem (see e.g. \cite{teufel}) we have that the
solution starts in the negative energy spectral  subspace of the free Dirac-Hamiltonian $D_0$ and ends in the
positive energy spectral subspace. Denoting with $P_0^+, P_0^-$ the corresponding spectral projectors we
formulate
\begin{cor}\label{psipropneu}(Adiabatic Pair Creation) For
$\psi_{s}^\varepsilon$ of Theorem \ref{onecrit} holds:\\ For all
$s<s_i$
\begin{equation}\label{onecriteq1}
\lim_{\varepsilon\to0}\langle\psi_{s}^\varepsilon,P_0^-\psi_{s}^\varepsilon\rangle=1\;.
\end{equation}
For all $s>s_f$
\begin{equation}\label{onecriteq2}
\lim_{\varepsilon\to0}\langle\psi_{s}^\varepsilon,P_0^+\psi_{s}^\varepsilon\rangle=1\;.
\end{equation}
\end{cor}
The proof of this consists in observing that (\ref{onecriteq1}) follows directly from the initial condition of
$\psi_{s}^\varepsilon$ (bound state in the gap) and the adiabatic theorem. For (\ref{onecriteq2}) one must apply
both the assertion of the theorem, which ensures that the scattering state is orthogonal to any bound state in
the gap, and thus the statement follows from the adiabatic theorem.

\section{Skeleton of the Proof and Content of
Sections}\label{skeleton} The proof of Theorem \ref{onecrit}
consists in controlling the propagation of $\psi^\varepsilon_s$. For
sufficiently small $\varepsilon$, $\psi^\varepsilon_s$ follows more
or less the bound states $\Phi_{s}$. Reaching the critical time
$s=0$ the bound state ``vanishes'' in the positive energy subspace
of the free Dirac Hamiltonian. One needs to show  that
$\psi^\varepsilon_s$ will stay there for all later times. Note that
we need to control the wavefunction evolution for a time dependent
hamiltonian.

The proof has thus naturally two parts:
\begin{enumerate}
\item[(1)] Show that a bound state with energy  in the energy gap
between $-1$ and $1$ reaches adiabatically the upper spectral edge
without ``injuries'' and ending up in $\mathcal{N}$. \item[(2)]
Show that any state in $\mathcal{N}$ scatters during the time in
which the potential stays overcritical.
\end{enumerate}

The proof of (1) is done in several  sections. Contrary to what
one might expect at first sight it is quite involved.
\begin{enumerate}
\item[(1.1)]  The problem is the possible degeneracy of the bound
states. We must show that the eigenspaces $\mathcal{N}_\mu$ of
$E\mu$ when $\mu$ goes to one converge to $\mathcal{N}$. That is
done in the first part of Section \ref{mu_derivative_of_phi}. {\it
Warning}: The proof is very long and tricky and aims at the
definition of an operator $R_\mu$, the resolvent of which maps a
state in  $\mathcal{N}$ to a state in $\mathcal{N}_\mu$. This map
will be used in the next step. \item[(1.2)] We need good control
of ``how a bound state converges''. We show that to every bound
state $\Phi\in \mathcal{N}$ exists a ``good sequence'' of bound
states $\Phi_\mu\in\mathcal{N}_\mu$ which are differentiable with
respect to $\mu$ when they approach $\Phi$. We use for that the
operator $R_\mu$ from above.  This is done in section \ref{51}.
 \item[(1.3)] The ``good sequences''  will be used in Section \ref{secco} to
 establish an adiabatic lemma without a gap, Lemma \ref{corvor0}.
 The proof  is a two scale argument. We first propagate
 adiabatically to times $s_0$ very close to $0$  and then by the
 uniform (in $\varepsilon$)
  estimate (\ref{uebersehen}) we can close the gap.
 This establishes the first part of the proof.
\end{enumerate}
The proof of (2) is naturally much more involved than that of (1).
\begin{enumerate}
\item[(2.1)] We wish to show that any  bound state at the spectral
edge scatters. What we need to establish is that the wavefunction
leaves the range of the scattering potential  sufficiently fast,
faster than $\varepsilon^{-1}$, the time after which the potential
is undercritical again. Such control is rather easy when the
potential is not depending on time, but here it depends on time:
How does one control the evolution of wavefunctions for time
dependent hamiltonians? The most direct and physical way is using
generalized eigenfunctions. The point is however that the
generalized eigenfunctions are bad near the spectral edge! The
essential question is: How bad? We recall here and rely heavily on
a result of \cite{picklneu,diss} on generalized eigenfunctions
near criticality. That is done in Section \ref{secgef}.
\item[(2.2)] The eigenfunctions allow us to control the
wavefunction evolution for potentials constant in time. Section
\ref{secpe} gives with Lemma \ref{propzeta} preliminary estimates.
(\ref{l1abschb2}) shows what we need to have when dealing with
generalized eigenfunctions namely an estimate in the sup-norm. The
proof involves tricky use of momentum cut-offs. Corollary
\ref{propestalt2} formulates then three estimates for the $L^2-$
norms. At first sight the third estimate ((iii) of Corollary
\ref{propestalt2}) seems the only relevant one, but the first two
are in fact needed for technical reasons later. (The technical
reason is that we must bring the ``static potential estimates'' in
contact with the true time evolution, i.e. with the non-static
potential). The proofs here are ``straightforward'' applications
of stationary phase methods, taking however the singular
eigenfunctions behavior into account by tricky cut-offs of small
momenta. The stationary phase application leading to good decay is
unfortunately lengthy, while not difficult or tricky. Therefore we
decided to shift that calculation to the Appendix.

 \item[(2.3)] In
section \ref{secpe1} the contact with the true time evolution is
made. First we consider the wavefunction evolution for ``short
times''.  Short means macroscopic times of order one. We introduce
the time $\sigma$ which may roughly be thought of as being the
first time at which $\mu$ reaches a maximum. Here we use the
estimate (iii) of Corollary \ref{propest}, which is the
translation of Corollary \ref{propestalt2} to the macroscopic time
scale. It is shown that most of the wavefunction will have left
the range of the potential by a macroscopic  time of order
$\varepsilon^{1/3}$, i.e. $\varepsilon^{-2/3}$ on the microscopic
time scale.  That proof uses Cook's argument in combination with
physical insights: We need to compare the true evolution until
time $s\le\sigma$ with a ``static potential'' evolution. The
potential will be ``frozen'' to the value it has at a time $s$.
There is a big error between the true and the auxiliary time
evolution. But in terms of the evolution of the relevant part of
the wavefunction the error is not so big, since most of the
wavefunction will have left the range of the potential. So the
error is transported to a region in space which we do not care so
much about. That idea is behind this part of the proof. Of course,
we must insure that for very long times, this error does not come
back! But that is done in the next step. \item[(2.4)]In Section
\ref{secpe2} we extend our result to all times. Most of the
wavefunction has left the range of the potential, we must insure
that it stays like that. Again we use Cook's method, but now we
freeze the potential at the value it has at time $\sigma$. In fact
we can chose here any value $s>0$ for which the potential is
overcritical. The physical idea is clear: Since most of the
wavefunction is outside of the range of the potential it moves
freely and the critical potential is roughly the zero potential.
To avoid problems arising from small $k$ values, we use density
arguments and cut off small momenta. It is here where Corollary
\ref{propest} (i) and (ii) come into play. \item[(2.5)] Section
\ref{endofproof} collects simply the results to establish the
second part of Theorem \ref{onecrit}, namely that the bound state
which reached the upper edge scatters.
\end{enumerate}

\section{Generalized Eigenfunctions}\label{secgef}

  Consider for $\mu\in \mathbb{R},$
$\mathbf{k}\in\mathbb{R}^3$ and $E_k=\sqrt{k^2+1}$ the bounded
classical solutions $\varphi_{\mu}(\mathbf{k},j,\mathbf{x})$
(generalized eigenfunctions (GEF)) of
\begin{equation}\label{dirac}
E_k\varphi_{\mu}(\mathbf{k},j)=D_{\mu}\varphi_{\mu}(\mathbf{k},j), \,\,j=1,2,3,4.
\end{equation}

\begin{lem} \label{properties}(GEF Properties)\\
Let $A$ satisfy condition \ref{cond} (i) and $\delta>0$ be such,
that $\mu=1$ is the only critical value in $[1,1+\delta]$. Then
\begin{description}
\item[(a)] there exist unique solutions
$\varphi_{\mu}(\mathbf{k},j,\cdot)$ of (\ref{dirac})

\item[(b)] for all $\mathbf{x} $  the functions
$\varphi^{j}(\mathbf{k},\mu,\mathbf{x})$ are infinitely often
continuously differentiable with respect to $k$ for $\mathbf{k}\ne
0\,.$

\item[(c)] The scattering system $(D_0,D_\mu=D_0+\mu A)$ is
asymptotically complete for any $\mu\in [1,1+\delta]$. In particular
the wave operator $\Omega_\mu^+$ defined via
$$\Omega_\mu^+\psi\equiv \lim_{t\to\infty}e^{iD_\mu t}e^{-iD_0t}\psi\;\;\;\text{for all }\psi\in L^2 $$
exists, is isometric and
$${\rm Ran}\; \Omega_\mu^+=\mathcal{H}^{cont}_\mu\;,$$
where $\mathcal{H}^{cont}_\mu$ is the spectral subspace of the absolutely continuous spectrum of $D_\mu$.

\item[(d)]
for $\mu\ne 1$ the solutions $\varphi_{\mu}(\mathbf{k},j,\cdot)$
    define a generalized Fourier transform, i.e. an isometry $\mathcal{F}:\mathcal{H}^{cont}_\mu\subset L^2(\mathbb{R}^3,\mathbb{C}^4)\rightarrow
    L^2((\mathbb{R}^3,\{1,2,3,4\}),\mathbb{C})$ by
\begin{equation}\label{her0}
\mathcal{F}_{\mu}(\psi)(\mathbf{k},j):=(2\pi)^{-\frac{3}{2}}\int\varphi_{\mu}(\mathbf{k},j,\mathbf{x})^*\psi(\mathbf{x})d^{3}x
\end{equation}
and
\begin{equation}\label{hin0}
    \psi(\mathbf{x})=\sum_{j=1}^{4}\int(2\pi)^{-\frac{3}{2}}
  \varphi_{\mu}(\mathbf{k},j,\mathbf{x})\mathcal{F}_{\mu}(\psi)(\mathbf{k},j)d^{3}k\;,
 \end{equation}
where the integrals are in the $l.i.m.$-sense (see e.g. \cite{reedsimon}). Furthermore Plancherel holds
\begin{eqnarray}\label{plancherel}
\langle\psi,\chi\rangle=\sum_{j=1}^4\int
\mathcal{F}_{\mu}(\psi)^*(\mathbf{k},j)\mathcal{F}_{\mu}(\chi)(\mathbf{k},j)d^3k
\end{eqnarray}
as well as Parseval
\begin{equation}\label{parseval}\|\psi\|=\sum_{j=1}^4\int
|\mathcal{F}_{\mu}(\psi)(\mathbf{k},j)|^2d^3k=:\|\mathcal{F}_{\mu}(\psi)\|\;.\end{equation}

\end{description}
\end{lem}
{\bf Proof:} (a) and (b) have been proven  in \cite{duerr} (see
Lemma 3.4. therein). Also (c) is not new, is is known to hold for
short range potentials (see for example Theorems 8.2, 8.3 and 8.20
in \cite{thaller}). (d) follows also from Lemma 3.4. in
\cite{duerr}, where it is shown that
$\mathcal{F}_{\mu}(\psi)(\mathbf{k},j)=\widehat{\psi}^{out}_\mu$
where the $\widehat{\cdot}$ stands for the (ordinary) Fourier
transform and $\psi^{out}_\mu$ is defined by
$\Omega_\mu^+\psi^{out}_\mu=\psi$.

Since $\Omega_\mu^+$ is isometric we have
\begin{eqnarray*}
\langle\psi,\chi\rangle&=&\langle\Omega_\mu^+\psi,\Omega_\mu^+\chi\rangle=\langle\psi^{out}_\mu,\chi^{out}_\mu\rangle
=\sum_{j=1}^4\int\widehat{\psi}^{out*}_\mu(\mathbf{k},j)\widehat{\chi}^{out}_\mu(\mathbf{k},j)d^3k
\\&=&\sum_{j=1}^4\int
\mathcal{F}_{\mu}(\psi)^*(\mathbf{k},j)\mathcal{F}_{\mu}(\chi)(\mathbf{k},j)d^3k\;,
\end{eqnarray*}
i.e. (\ref{plancherel}) and (\ref{parseval}) hold. $\square$

As we  deal with a time dependent external field which grows from
under-criticality to over-criticality in the course of which we need
very good control on the evolution of the wavefunction, we need {\em
uniform} estimates on the generalized eigenfunctions of the operator
$D_0+\mu A$. Uniform estimates  have not  been given before. It is
known, that for critical potentials the generalized eigenfunctions
diverge for $k\to0$ \cite{jensen}, but that is not sufficient to
control the propagation. What is sufficient are estimates on the
$L^\infty$-norm of the generalized eigenfunctions of $D_0+\mu A$ and
their $k$-derivatives {\em uniform } in $\mathbf{k}$ and {\em
uniform} in $\mu\in[1-\delta,1+\delta]$ for some $\delta>0$. The
uniform estimates we need are provided in \cite{picklneu}. We cite
the the following crucial Corollary 7.5. in \cite{picklneu}.

\begin{thm}\label{gefthm}(Upper Bound for the Sup-Norm of the GEF of Lemma \ref{properties})
There exist $\delta>0$, constants $\nu_l$, $1\leq l\leq n$ and a
constant $c>0$ so that the following holds: For the $m^{th}$
derivative $\varphi_\mu^{(m)}:=\partial_k^m\varphi_\mu$,
$m\in{\mathbb{N}}_0,$ there exist constants $C_m$  so that for
every $\mathbf{k}\in\mathbb{R}^3$ and for every
$\mu\in[1-\delta,1+\delta]$
\begin{equation}\label{resultat2thm}
\|(1+x)^{-m}\varphi_\mu^{(m)}(\mathbf{k},j,\cdot) \|_\infty<C_m\left(k^{-m}+\left|\sum_{l=1}^n\frac{k}{
 |\mu-1- \nu_l k^2|
+ck^3}\right|^{m+1}\right)\;.
\end{equation}
Furthermore there exist
$\Omega_\mu(\mathbf{k},j,\cdot)\in\mathcal{N}$ and  $C$ uniform in
$\mathbf{k}\in\mathbb{R}^3$ and $\mu\in[1-\delta,1+\delta]$ so
that
\begin{equation}\label{resultat2}
\|\varphi_\mu(\mathbf{k},j,\cdot)-\Omega_\mu(\mathbf{k},j,\cdot)\|_\infty<C\;.
\end{equation}
\end{thm}

\section{Propagation Estimates}\label{secpe}

In this section we want to apply Theorem \ref{gefthm} to get
estimates on the time propagation of wave-functions for the static
Dirac Hamiltonian $D_\mu$ uniform in $\mu\in[1-\delta,1+\delta]$.

Under Condition \ref{cond}  (see e.g. \cite{thaller})  $D_\mu$
generates a unitary time evolution denoted by $V_\mu(t,0)$, i.e.
\begin{equation}\label{V}
i\partial_t V_\mu(t,0)= D_\mu V_\mu(t,0)\,,
\end{equation}
which applied to eigenfunctions reads
\begin{equation}\label{wichtigeformel}
V_\mu(t,0)\varphi_\mu(\mathbf{k},j,\mathbf{x})=e^{-iE_k
t}\varphi_\mu(\mathbf{k},j,\mathbf{x})\,.\end{equation} This formula
explains the role of the generalized eigenfunctions and it gives us
the most direct control on the evolution of wavefunctions by
expanding the wavefunction into generalized eigenfunctions. The
estimates we are after are such that we can control the wavefunction
evolution of the bound states in $\mathcal{N}$ during
overcriticality. The bound states must decay fast enough (i.e.
scatter fast enough) so that they are outside of the range
$\mathcal{S}_A$ of the potential before the potential becomes
undercritical again. The naive picture of scattering theory suggests
that the Fourier transform (given by plane waves) of the state
governs the spreading. But we are here in a delicate situation
analogous to resonant behavior. The generalized eigenfunctions are
not at all like plane waves as we see from (\ref{resultat2thm}) and
we must control the spreading given by such badly behaved
eigenfunctions. We need to separate very very slow spreadings of the
wavefunction (whose contribution will be hopefully negligible
because of small probability) from the moderately fast spreading
which make the state scatter. The borderline will be given by the
$k$-value of the ``resonance'', i.e., where $\mu-1 \approx \nu_l
k^2$ (c.f. (\ref{resultat2thm})). $\mu$ should be thought of being
only slightly bigger than $1$, because that is the dangerous regime,
the regime where $k$ is small with large probability. For technical
reasons we also separate very large momenta. Thus we will give
propagation estimates for a wavefunction separating large,
intermediate and small momenta, using the mollifier
$\widehat{\rho}_\kappa\in C^\infty$ given by
\begin{equation}\label{rhodef2}
\widehat{\rho}(\mathbf{k}):=\left\{%
\begin{array}{ll}
    0, & \hbox{for $k\leq1$,} \\
    1, & \hbox{for $k\geq2$;} \\
\end{array}%
\right.
\end{equation}
and for $\kappa>0$ we define
\begin{equation}\label{defrho2}
\widehat{\rho}_{\kappa}(\mathbf{k}):=\widehat{\rho}\left(\frac{\mathbf{k}}{\kappa}\right)
\end{equation}
and the operator
\begin{equation}\label{rhohut}\rho_{\kappa,\mu}:=\mathcal{F}_\mu^{-1}\widehat{\rho}_\kappa\mathcal{F}_\mu\;.\end{equation}
Note that $[D_\mu,\rho_{\kappa,\mu}]=0$.

\begin{lem}\label{propzeta}(Cutoff and propagation estimates - stationary case)\\
Let $\delta>0$ be such that there is no bound state of $D_\mu$ for
$\mu\in(1,1+\delta]$, i.e. $\mathcal{H}^{ac}(D\mu)=L^2$. Let
$\mathcal{S}\subset\mathbb{R}^3$ be compact. For any $\chi\in L^2$
with ${\rm supp}\chi\subset\mathcal{S}$ and any
$0<\underline{\kappa}<1$ we have that for all $\mu\in(1,1+\delta]$
and all $0\le t<\infty$

\begin{eqnarray}\label{propzetaeq1}
\|(1-\rho_{\underline{\kappa},\mu})\chi\|
\leq C \underline{\kappa}^{\frac{3}{2}}\sup_{k<2\underline{\kappa}}\left(1+\left|\sum_{l=1}^n\frac{k}{
 |\mu-1- \nu_l k^2|
+ck^3}\right|\right)
\end{eqnarray}
and
\begin{eqnarray}\label{propzetaeq2}
\| V_\mu(t,0)(1-\rho_{\underline{\kappa},\mu})\chi\|_\infty
\leq C\underline{\kappa}^{3}\sup_{k<2\underline{\kappa}}\left(1+\left|\sum_{l=1}^n\frac{k}{
 |\mu-1- \nu_l k^2|
+ck^3}\right|\right)^2\,.
\end{eqnarray}
Furthermore let  $\overline{\kappa}<\infty$. For all $m\in\mathbb{N}_0$ there exists $C_m<\infty$ such that
\begin{eqnarray}\label{l1abschb2}
&&\|
\mathds{1}_\mathcal{S}V_\mu(t,0)\rho_{\underline{\kappa},\mu}(1-\rho_{\overline{\kappa},\mu})\chi\|_\infty
\\\nonumber&&\hspace{1cm}<
\overline{\kappa}^3t^{-m}C_n\underline{\kappa}^{-4}\left(\underline{\kappa}^{-2}+\sup_{2\overline{\kappa}>k>\underline{\kappa}}\left|\sum_{l=1}^n\frac{1}{
 |\mu-1- \nu_l k^2|
+ck^3}\right|\right)^{m}
\end{eqnarray}
and
\begin{eqnarray}\label{l1abschc}
\| \rho_{\underline{\kappa},\mu}\rho_{\overline{\kappa},\mu}\chi\|
&\leq&\frac{\|D_\mu\chi\|}{\overline{\kappa}} \;.
\end{eqnarray}
\end{lem}

\noindent\textbf{Proof:} We start with (\ref{propzetaeq1}). Let $\chi$ be as in the Lemma. Then by
(\ref{rhohut}) and (\ref{parseval})
\begin{eqnarray}
\label{esteta}
  \|(1-\rho_{\underline{\kappa},\mu})\chi\|=\|(1-\widehat{\rho}_{\underline{\kappa}})\mathcal{F}_\mu(\chi)\|
  \leq4\sup_{j,k<2\underline{\kappa}}\{\mid\mathcal{F}_\mu(\chi)(\mathbf{k},j)\mid\}\|1-\widehat{\rho}_{\underline{\kappa}}\|
\end{eqnarray}
By (\ref{defrho2}) and (\ref{rhodef2})
\begin{eqnarray}\label{substk0raus}
\|\widehat{\rho}_{\underline{\kappa}}-1\|=\underline{\kappa}^{\frac{3}{2}}\left(\int\mid\widehat{\rho}(p)-1\mid^2d^3p\right)^{\frac{1}{2}}\leq
C\underline{\kappa}^{3/2}\;.
\end{eqnarray}
Furthermore
\begin{eqnarray}\label{obenrein}
\sup_{j,k<2\underline{\kappa}}\{\mid\mathcal{F}_\mu(\chi)(\mathbf{k},j)\mid\}&\leq&
\sup_{j,k<2\underline{\kappa}}\{
\int(2\pi)^{-\frac{3}{2}}\mid\overline{\varphi}_{\mu}(\mathbf{k},j,\mathbf{x})\chi(\mathbf{x})\mid
d^{3}x \}
\nonumber\\&\leq&
\sup_{j,k<2\underline{\kappa}}\{\|\varphi_{\mu}(\mathbf{k},j,\cdot)\|_\infty\}
(2\pi)^{-\frac{3}{2}}\|\chi\|_1\;,
\end{eqnarray}
where by Schwartz
\begin{eqnarray}\label{l1l2}
\|\chi\|_1&=& \left|\int\mid\chi(\mathbf{x})\mid d^{3}x\right|
\nonumber\\&=&\left|\int\mathds{1}_{\mathcal{S}}\mid\chi(\mathbf{x})\mid d^{3}x\right|
\nonumber\\&\leq&\|\chi(\mathbf{x})\|\sqrt{\mid\mathcal{S}\mid}\leq
C\,.
\end{eqnarray}
For $\sup_{j,k<2\underline{\kappa}}\{\|\varphi_{\mu}(\mathbf{k},j,\cdot)\|_\infty\}$ we have by Theorem
\ref{gefthm} (\ref{resultat2thm}) for $m=0$ that
\begin{equation}\label{phiklein}
\sup_{j,k<2\underline{\kappa}}\{\|\varphi_{\mu}(\mathbf{k},j,\cdot)\|_\infty\}\leq
C\sup_{k<2\underline{\kappa}}\left(1+\left|\sum_{l=1}^n\frac{k}{
 |\mu-1- \nu_l k^2|
+ck^3}\right|\right)\;.
\end{equation}
Hence for (\ref{obenrein}) with (\ref{l1l2})
\begin{equation}\label{supchi}
\sup_{j,k<2\underline{\kappa}}\{\mid\mathcal{F}_\mu(\chi)(\mathbf{k},j)\mid\}\leq
C\sup_{k<2\underline{\kappa}}\left(1+\left|\sum_{l=1}^n\frac{k}{
 |\mu-1- \nu_l k^2|
+ck^3}\right|\right)\;.
\end{equation}
This  with (\ref{substk0raus}) in (\ref{esteta})  yields
(\ref{propzetaeq1}).

Next we establish (\ref{propzetaeq2}). Observing the definitions
\begin{eqnarray*}
&&\| V_\mu(t,0)(1-\rho_{\underline{\kappa},\mu})\chi\|_\infty
\\&&\hspace{0.5cm}\leq\nonumber\sum_{j=1}^4\left\|\int(2\pi)^{-\frac{3}{2}}\mid
V_\mu(t,0)\varphi_{\mu}(\mathbf{k},j,\mathbf{x})
\mathcal{F}_\mu(\chi)(\mathbf{k},j)(1-\widehat{\rho}_{\underline{\kappa}}(\mathbf{k}))\mid d^3k\right\|_\infty
\\&&\hspace{0.5cm}=\nonumber\sum_{j=1}^4\left\|\int(2\pi)^{-\frac{3}{2}}\mid e^{-iE_kt}\varphi_{\mu}(\mathbf{k},j,\mathbf{x})
\mathcal{F}_\mu(\chi)(\mathbf{k},j)(1-\widehat{\rho}_{\underline{\kappa}}(\mathbf{k}))\mid d^3k\right\|_\infty
\\&&\hspace{0.5cm}\leq4\sup_{j,k<2\underline{\kappa}}\{\|\varphi_\mu(\mathbf{k},j,\cdot)\|_\infty\}\sup_{j,k<2\underline{\kappa}}\{\mid\mathcal{F}_\mu(\chi)(\mathbf{k},j)\mid\}
\|1-\widehat{\rho}_{\underline{\kappa}}\|_1\;.
\end{eqnarray*}
Similarly as in (\ref{substk0raus}) we find that
$\|1-\widehat{\rho}_{\underline{\kappa}}\|_1<C\underline{\kappa}^3$ and with (\ref{phiklein}) and
(\ref{supchi})) we get (\ref{propzetaeq2}).

We now turn to (\ref{l1abschb2}). As above
\begin{eqnarray}
&&\label{produkte}\hspace{-0.5cm}V_\mu(t,0)\rho_{\underline{\kappa},\mu}(1-\rho_{\overline{\kappa},\mu})\chi(\mathbf{x})
\nonumber\\&&=\sum_{j=1}^{4}\int(2\pi)^{-\frac{3}{2}}
  V_\mu(t,0)\varphi_{\mu}(\mathbf{k},j,\mathbf{x})\widehat{\rho}_{\underline{\kappa}}(1-\widehat{\rho}_{\overline{\kappa}})\mathcal{F}_\mu(\chi)(\mathbf{k},j)d^{3}k
\nonumber\\&&=\sum_{j=1}^{4}\int(2\pi)^{-\frac{3}{2}}
 \exp\left(-itE_k\right) \varphi_{\mu}(\mathbf{k},j,\mathbf{x})
 \widehat{\rho}_{\underline{\kappa}}(1-\widehat{\rho}_{\overline{\kappa}})\mathcal{F}_\mu(\chi)(\mathbf{k},j)d^{3}k\;.
\end{eqnarray}
It is for this term that we need the behavior of the derivatives of
the eigenfunctions (c.f. (\ref{resultat2thm})). We shall use a
stationary phase method, using $\frac{i
E_k}{kt}\partial_k\exp\left(-i tE_k\right)=\exp\left(-i
tE_k\right)$.
 The rigorous estimate of this formula is based on a simple straightforward computation which is done in the
appendix. We shall only describe here in a heuristic manner how the estimate comes about.

First we recall (\ref{resultat2thm})
\begin{equation}\label{31}\|(1+x)^{-m}\varphi_\mu^{(m)}(\mathbf{k},j,\cdot) \|_\infty\leq
C_m\left(k^{-m}+\left|\sum_{l=1}^n\frac{k}{
 |\mu-1- \nu_l k^2|
+ck^3}\right|^{m+1}\right)\;.\end{equation} This enters also in the $k$-derivatives of $\mathcal{F}_\mu(\chi)$.
Since $\chi$ has compact support in $\mathcal{S}$ we obtain
\begin{eqnarray}\label{32}
|\partial_k^m\mathcal{F}_\mu(\chi)(\mathbf{k},j)|&=&|\partial_k^m\langle\varphi_\mu(\mathbf{k},j,\cdot),\chi\rangle|
\nonumber\\&\leq&
C_m\|(1+x)^{-m}\varphi_\mu^{(m)}(\mathbf{k},j,\cdot)\|_\infty\|(1+x)^m\chi\|_1
\nonumber\\&\leq& C
C_m\|(1+x)^{-m}\varphi_\mu^{(m)}(\mathbf{k},j,\cdot)\|_\infty
\end{eqnarray}
where in the last step we followed (\ref{l1l2}). Doing the partial integration we need to apply the operator
$\partial_k \frac{E_k}{k}=\frac{1}{k^2}+\frac{E_k}{k}\partial_k$. Relevant to us is only the ``small''
$k$-behavior ($k\geq\underline{\kappa}$), i.e. we need to count the inverse powers of $k$. In that sense
$\partial_k \frac{E_k}{k}\sim\frac{1}{k^2}+\frac{1}{k}\partial_k$. Further observe that the relevant term in
(\ref{produkte}) to which $\partial_k \frac{E_k}{k}$ is applied is the product
$\varphi_{\mu}(\mathbf{k},j,\mathbf{x})
\widehat{\rho}_{\underline{\kappa}}(1-\widehat{\rho}_{\overline{\kappa}})\mathcal{F}_\mu(\chi)(\mathbf{k},j)$.
But $\partial_k \widehat{\rho}_{\underline{\kappa}}\sim\frac{1}{\underline{\kappa}}\sim\frac{1}{k}$, while
\begin{eqnarray*}
\partial_k
\varphi^{(m)}_{\mu}(\mathbf{k},j,\mathbf{x})&=&\varphi^{(m+1)}_{\mu}(\mathbf{k},j,\mathbf{x})\\\nonumber&\approx&(1+x)
\varphi^{(m)}_{\mu}(\mathbf{k},j,\mathbf{x})\left(k^{-1}+\left|\sum_{l=1}^n\frac{k}{
 |\mu-1- \nu_l k^2|
+ck^3}\right|\right)\;. \end{eqnarray*}Likewise for $\mathcal{F}_\mu(\chi)$.  Since we are interested in the
supremum over the compact set $\mathcal{S}$ the factor $(1+x)$ can be estimated by a constant. The upshot is
then that the contribution of the terms is essentially the $m$-th power of
\begin{eqnarray*}\partial_k \frac{E_k}{k}&\approx& \frac{1}{k^2}+\frac{1}{k}\left(k^{-1}+\left|\sum_{l=1}^n\frac{k}{
 |\mu-1- \nu_l k^2|
+ck^3}\right|\right)
\\&\approx&\frac{1}{k^2}+\left|\sum_{l=1}^n\frac{1}{
 |\mu-1- \nu_l k^2|
+ck^3}\right|\end{eqnarray*}  multiplied by the product
$\varphi_{\mu}\mathcal{F}_\mu(\chi)$,  yielding roughly
$$\left(\frac{1}{k^2}+\left|\sum_{l=1}^n\frac{1}{
 |\mu-1- \nu_l k^2|
+ck^3}\right|\right)^{m}\left(1+\left|\sum_{l=1}^n\frac{k}{
 |\mu-1- \nu_l k^2|
+ck^3}\right|\right)^2\;.$$ The second factor can be bounded by $c^{-2}k^{-4}\leq C \underline{\kappa}^{-4}$.
This gives with the volume factor $\overline{\kappa}^3$ the right hand side of (\ref{l1abschb2}).

Finally we establish (\ref{l1abschc}). By (\ref{parseval}) and using $E_k^2\geq k^2$
\begin{eqnarray*}
\|\rho_{\underline{\kappa},\mu}\rho_{\overline{\kappa},\mu}\chi\|^2
&\leq&\int_{k>\overline{\kappa}}|\mathcal{F}_\mu(\chi)|^2d^3k
\leq\int_{k>\overline{\kappa}}E_{k}^{-2}\mathcal{F}_\mu(D_\mu\chi)^*\mathcal{F}_\mu(D_\mu\chi)d^3k
\\&\leq&\frac1{
\overline{\kappa}^{\,2}}\int_{k>\overline{\kappa}}\mathcal{F}_\mu(D_\mu\chi)^*\mathcal{F}_\mu(D_\mu\chi)d^3k
=\frac{\|D_\mu\chi\|^2 }{\overline{\kappa}^{\,2}}\;.
\end{eqnarray*}

$\square$.

The results of Lemma \ref{propzeta} can now be used to estimate the decay behavior of any compactly supported
wavefunction $\chi\in L^2$

\begin{cor}\label{propestalt2}(Propagation Estimate - stationary case)\\
Let  $\mathcal{S}\subset\mathbb{R}^3$ be compact. There exists a $\delta>0$ (possibly smaller than the $\delta$
of Lemma \ref{propzeta}) such that for all $\widetilde{m}\in\mathbb{N}$ and for all $0<\xi<1$ exist constants
$C_{\xi,\widetilde{m}}$ and $C_\xi$ such that for all $\mu\in(1,1+\delta]$, all
$t>(\mu-1)^{-\frac{3}{2(1-\xi)}}$ and all $\chi\in L^2$ with ${\rm supp}\chi\subset\mathcal{S}$ the following
holds:
\begin{itemize}
\item[(i)] Let $V_{\mu}$ be defined by (\ref{V}), then for $\underline{\kappa}=t^{-\frac{1}{2}(1-\xi)}$ and for
all $w\geq t$
\begin{eqnarray*}
\|\mathds{1}_\mathcal{S}
V_{\mu}(w,0)\rho_{\underline{\kappa},\mu}\chi\|
&\leq&C_{\xi,\widetilde{m}}(\|D_\mu\chi\|)w^{-\widetilde{m}}\;,
\end{eqnarray*}

\item[(ii)] for $\underline{\kappa}=t^{-\frac{1}{2}(1-\xi)}$
$$\|(1-\rho_{\underline{\kappa},\mu})\chi\|\leq C_{\xi}t^{-\frac{3}{4}(1-\xi)}(\mu-1)^{-\frac{1}{4}}\|\chi\|\;,$$

\item[(iii)]
\begin{eqnarray*}
\|\mathds{1}_\mathcal{S} V_{\mu}(t,0)\chi\|
&\leq&C_\xi(\|\chi\|+\|D_\mu\chi\|)|\mu-1|^{-\frac{1}{2}}t^{-\frac{3}{2}(1-\xi)}\;.
\end{eqnarray*}
\end{itemize}
\end{cor}
We note that we want to have good decay estimates, so $\xi$ should be thought of being small. We also wish to
remark that we shall need both estimates (i) and (ii) as well as (iii). (iii) is better than (i) and (ii)
together but it is not suitable for ``density arguments'' which we shall use later on when we compare the true
time evolution with the $V_\mu$-evolution.

\noindent{\bf Proof:}  By linearity we may assume $\|\chi\|=1$. We start with (ii). With our choice
$\underline{\kappa}=t^{-\frac{1}{2}(1-\xi)}$ we obtain in view of (\ref{propzetaeq1})
\begin{eqnarray*}
\|(1-\rho_{\underline{\kappa},\mu})\chi\|\leq
Ct^{-\frac{3}{4}(1-\xi)}\sup_{k<2\underline{\kappa}}\left(1+\sum_{l=1}^n\frac{k}{
 |\mu-1- \nu_l k^2|
+ck^3}\right)\;.
\end{eqnarray*}
Since by assumption $t>(\mu-1)^{-\frac{3}{2(1-\xi)}}$ we have $\underline{\kappa}^2<(\mu-1)^{3/2}\ll (\mu-1)$
and hence  for $\delta$ small enough  we are below the resonant $k$-values, i.e. $\inf_l|\mu-1-\nu_lk^2|\geq
(\mu-1)/2$ and thus the supremum of the bracket term is less than $\left(1+C\underline{\kappa}/(\mu-1)\right)$.
Thus
\begin{eqnarray*}
\|(1-\rho_{\underline{\kappa},\mu})\chi\|&\leq&
Ct^{-\frac{3}{4}(1-\xi)}\left(1+\frac{C\underline{\kappa}}{\mu-1}\right)
\\&\leq& Ct^{-\frac{3}{4}(1-\xi)}\left(1+\frac{C(\mu-1)^{3/4}}{\mu-1}\right)
\\&\leq&Ct^{-\frac{3}{4}(1-\xi)}(\mu-1)^{-1/4}\;,
\end{eqnarray*}
which establishes (ii).

We now prove (i). Using
\begin{equation}\label{kompakt}\|\mathds{1}_\mathcal{S}\psi\|\leq\|\mathds{1}_\mathcal{S}\|\;\|\mathds{1}_\mathcal{S}\psi\|_\infty\leq C\|\mathds{1}_\mathcal{S}\psi\|_\infty\end{equation}
we have with the high momentum cutoff $\overline{\kappa}$ to be
specified below
\begin{eqnarray}\label{dreiterme}
&&\|\mathds{1}_\mathcal{S}\nonumber
V_{\mu}(w,0)\rho_{\underline{\kappa},\mu}\chi\|\\&&\leq
\|\mathds{1}_\mathcal{S}\nonumber
V_{\mu}(w,0)\rho_{\underline{\kappa},\mu}(1-\rho_{\overline{\kappa},\mu})\chi\|+\|\mathds{1}_\mathcal{S}\nonumber
V_{\mu}(w,0)\rho_{\underline{\kappa},\mu}\rho_{\overline{\kappa},\mu}\chi\|
\nonumber\\&&\leq
C\|\mathds{1}_\mathcal{S}V_{\mu}(w,0)\rho_{\underline{\kappa},\mu}(1-\rho_{\overline{\kappa},\mu})\chi\|_\infty
+\|\rho_{\underline{\kappa},\mu}\rho_{\overline{\kappa},\mu}\chi\|\;.
\end{eqnarray}
Let $w\geq t> (\mu-1)^{-\frac{3}{2(1-\xi)}}$. $\overline{\kappa}$
will be chosen large enough, so that the first term encompasses the
resonant regime ($\nu_l k^2\approx\mu-1$). We obtain with
(\ref{l1abschb2}) using $w\geq t> (\mu-1)^{-\frac{3}{2(1-\xi)}}$ and
$\underline{\kappa}=t^{-\frac{1}{2}(1-\xi)}$
\begin{eqnarray*}
\|
\mathds{1}_\mathcal{S}V_\mu(w,0)\rho_{\underline{\kappa},\mu}(1-\rho_{\overline{\kappa},\mu})\chi\|_\infty
\nonumber&\leq&CC_m
\overline{\kappa}^3w^{-m}t^{2-2\xi}\left(t^{1-\xi}+C(\mu-1)^{-\frac{3}{2}}\right)^m
\\&\leq&CC_m \overline{\kappa}^3w^{-m}\left(Ct^{1-\xi}\right)^{m+2}
\nonumber\\&\leq&\widetilde{C}_m
\overline{\kappa}^3w^{-m\xi+2-2\xi}\;.
\end{eqnarray*}
For the second term in (\ref{dreiterme}) we get with
(\ref{l1abschc})
\begin{eqnarray*}\|\rho_{\underline{\kappa},\mu}\rho_{\overline{\kappa},\mu}\chi\|
&\leq&\frac{\|D_\mu\chi\|}{\overline{\kappa}}\;.
\end{eqnarray*}
Hence for (\ref{dreiterme}) we obtain
\begin{eqnarray*}
\|\mathds{1}_\mathcal{S}\nonumber
V_{\mu}(w,0)\rho_{\underline{\kappa},\mu}\chi\|
\leq\widetilde{C}_m \overline{\kappa}^3w^{-m\xi+2-2\xi}+\frac{\|D_\mu\chi\|}{\overline{\kappa}}\;.
\end{eqnarray*}
Choosing $\overline{\kappa}:=w^{m\xi/4}$ we find the bound
\begin{eqnarray*}
\|\mathds{1}_\mathcal{S}\nonumber
V_{\mu}(w,0)\rho_{\underline{\kappa},\mu}\chi\|\leq\widetilde{C}_m\|D_\mu\chi\| w^{-\frac{m\xi}{4}+2-2\xi}\;.
\end{eqnarray*}
Choosing $m$ such that  $-m\xi/4+2-2\xi>\widetilde{m}$ (i) follows.

Next we prove (iii). By (\ref{kompakt})
\begin{eqnarray*}
\|\mathds{1}_\mathcal{S}
V_{\mu}(t,0)\chi\|&\leq&\|\mathds{1}_\mathcal{S}
V_{\mu}(t,0)\rho_{\underline{\kappa},\mu}\chi\|+\|\mathds{1}_\mathcal{S}
V_{\mu}(t,0)(1-\rho_{\underline{\kappa},\mu})\chi\|_\infty
\\&\leq&\|\mathds{1}_\mathcal{S}
V_{\mu}(t,0)\rho_{\underline{\kappa},\mu}\chi\|+\|
V_{\mu}(t,0)(1-\rho_{\underline{\kappa},\mu})\chi\|_\infty \;.
\end{eqnarray*}
For the first summand use (i) with $w=t$ and $\widetilde{m}=2$. For the second summand use (\ref{propzetaeq2})
with $\underline{\kappa}=t^{-\frac{1}{2}(1-\xi)}$ to obtain
\begin{eqnarray*}
\|V_{\mu}(t,0)(1-\rho_{\underline{\kappa},\mu})\chi\|_\infty&\leq&
Ct^{-\frac{3}{2}(1-\xi)}\sup_{k<2\underline{\kappa}}\left(1+\sum_{l=1}^n\frac{k}{
 |\mu-1- \nu_l k^2|
+ck^3}\right)^2
\\&\leq&C|\mu-1|^{-\frac{1}{2}}t^{-\frac{3}{2}(1-\xi)}
\;,
\end{eqnarray*}
where the bound comes as in the proof of (ii) with the only difference being that we now have the
square.

$\square$
\section{$\mu$-Convergence of the Eigenspaces}\label{mu_derivative_of_phi}
Let $[\mu_B,1]$ be the be the interval of parameters  for which
bound states for
$$D_{\mu}\Phi_{\mu}:=(D_0+\mu A)\Phi_{\mu}=E_\mu\Phi_{\mu}$$
exist. In the course of this paper we shall adjust $\mu_B<1$
according to our needs.  Note that $E_\mu\in[-1,1]$. Let
$\mathcal{N}_\mu$ denote the eigenspace of $E_\mu$. In \cite{klaus}
it is shown that Condition \ref{cond} (i) implies that there exist
constants $\mu_B<1$, $\underline{C}>0$ and $\overline{C}>0$ such
that

\begin{equation}\label{diveproperly}
\underline{C}<\partial_\mu E_\mu<\overline{C}
\end{equation}
for all $\mu_B\leq \mu\leq 1$.

\begin{lem}\label{sequenz}Let $A$ satisfy Condition \ref{cond}. Let $P_{\mathcal{N}}$ be the projector onto
$\mathcal{N}$.
\begin{itemize}
\item[(i)] For any sequence $(\Phi_\mu)_\mu, \,\, \Phi_\mu\in\mathcal{N}_\mu, \,\,\|\Phi_\mu\|=1$,
\begin{equation}\label{wahr}
\lim_{\mu\to1}\|P_{\mathcal{N}}\Phi_\mu\|=1\;.
\end{equation}
\item[(ii)] For $\mu_B$ close enough to $1$
$$\dim \mathcal{N}=\dim \mathcal{N}_\mu$$
for all $\mu\in[\mu_B,1)$.
\end{itemize}
\end{lem}
{\bf Proof:} We shall present a proof which prepares notation and
results which we shall need later on for the control of the
$\mu$-derivative of the bound states. Therefore there might be
shorter proofs of the lemma. Our aim is to define first an operator
$R_\mu$ the resolvent of which maps states in $\mathcal{N}$ to
states in $\mathcal{N}_\mu$. The resolvent will be written as
geometric series and we need good control on the norm of $R_\mu$. So
before we prove the lemma we shall be concerned with $R_\mu$ the
upshot of which is Corollary \ref{thecorollary}. Having that the
actual proof of the lemma is short.

  Let
$\Phi_\mu\in\mathcal{N}_\mu$ be normalized, let
$\Phi\in\mathcal{N}$ be normalized and such that
$\Phi_\mu\in(\mathcal{N}\backslash\Phi)^\bot$ (where
$(\mathcal{N}\backslash\Phi)^\bot$ is the orthogonal complement of
$\mathcal{N}\backslash\Phi$). Such a  $\Phi$ exists: If
$P_\mathcal{N}\Phi_\mu\neq 0$ then
$\Phi=\frac{P_\mathcal{N}\Phi_\mu}{\|P_\mathcal{N}\Phi_\mu\|}$, if
$P_\mathcal{N}\Phi_\mu=0$ one can choose for $\Phi$ any normalized
element of $\mathcal{N}$. Hence in general $\Phi$ depends on the
choice of $\Phi_\mu$ but we shall refrain from indicating that
further to not overburden the proof with notation.

With
\begin{equation}\label{skpmitphi}(
D_{\mu}-E_\mu)\Phi_\mu=0\,
\end{equation}
 and \begin{equation}\label{ddiff}
D_\mu-D_\nu=(\mu-\nu)A\end{equation} we have
\begin{eqnarray*}
0&=&\langle (  D_{\mu}-E_\mu)\Phi_\mu,\Phi\rangle
\\&=&\langle (
D_{1}-E_\mu)\Phi_\mu,\Phi\rangle+\langle
(\mu-1)A\Phi_\mu,\Phi\rangle
\\&=&(
E_1-E_\mu)\langle \Phi_\mu,\Phi\rangle+\langle
(\mu-1)A\Phi_\mu,\Phi\rangle
%
\;.
\end{eqnarray*}
Thus
\begin{eqnarray*}
(E_1-E_\mu)\langle\Phi_\mu,\Phi\rangle\Phi&=& (1-\mu)\langle
A\Phi_\mu,\Phi\rangle\Phi
\end{eqnarray*}
or
$$(D_1-E_\mu)\langle\Phi_\mu,\Phi\rangle\Phi= (1-\mu)\langle A\Phi_\mu,\Phi\rangle\Phi\;.$$
Since for $\mu\in[\mu_B,1)$ $E_\mu$  is in the resolvent set of
$D_1$
\begin{eqnarray}\label{phidminuse}
\langle\Phi_\mu,\Phi\rangle\Phi=(1-\mu)\langle
A\Phi_\mu,\Phi\rangle\left(D_1-E_\mu\right)^{-1}\Phi\;.
\end{eqnarray}
On the other hand, by (\ref{skpmitphi}) and (\ref{ddiff}),
\begin{equation}\label{schoenermachen}\left(D_1-E_\mu\right)\Phi_\mu=(1-\mu)A\Phi_\mu\;,\end{equation}
i.e.
\begin{equation}\label{schoenermachenb}0=\Phi_\mu-(1-\mu)\left(D_1-E_\mu\right)^{-1}A\Phi_\mu\;.\end{equation}
Adding (\ref{schoenermachenb}) to (\ref{phidminuse}) yields
\begin{eqnarray*}
\langle\Phi_\mu,\Phi\rangle\Phi&=&\Phi_\mu-(1-\mu)\left(
D_{1}-E_\mu\right)^{-1}\left(A\Phi_\mu-\langle
A\Phi_\mu,\Phi\rangle\Phi\right)\;.
\end{eqnarray*}
$(\mathcal{N}\backslash\Phi)^\bot$ is an invariant  subspace of
$D_1$ and since $\Phi_\mu\in(\mathcal{N}\backslash\Phi)^\bot$ the
left hand side of (\ref{schoenermachen}) is in
$(\mathcal{N}\backslash\Phi)^\bot$ and hence
$A\Phi_\mu\in(\mathcal{N}\backslash\Phi)^\bot$. Therefore
 $\langle
A\Phi_\mu,\Phi\rangle\Phi=P_\mathcal{N} (A\Phi_\mu)$. Writing
$P_{\mathcal{N}^\bot}$ for the projector on $\mathcal{N}^\bot$ ---
the orthogonal complement of $\mathcal{N}$
--- it follows that
\begin{eqnarray}\label{adding}
\langle\Phi_\mu,\Phi\rangle\Phi&=&\Phi_\mu-(1-\mu)\left(
D_{1}-E_\mu\right)^{-1}\left( A\Phi_\mu-P_\mathcal{N}
(A\Phi_\mu)\right)
\nonumber\\&=&\Phi_\mu-(1-\mu)\left(
D_{1}-E_\mu\right)^{-1}P_{\mathcal{N}^\bot} (A\Phi_\mu)\;.
\end{eqnarray}
Note that the argument of $(D_1-E_\mu)^{-1}$ is now a vector orthogonal to $\mathcal{N}$. Therefore the term has
good chances of being controllable. Again observing $[D_1,P_{\mathcal{N}^\bot}]=0$, define
$$R_\mu:(\mathcal{N}\backslash\Phi)^\bot\rightarrow\mathcal{N}^\bot$$
by
\begin{equation}\label{defrop}
R_\mu\chi:=\left(D_{1}-E_\mu\right)^{-1}P_{\mathcal{N}^\bot}\left(A\chi\right)\;.
\end{equation}
Then by (\ref{adding})
\begin{eqnarray}\label{rreihe}
\langle\Phi_\mu,\Phi\rangle\Phi&=&\Phi_\mu-(1-\mu)R_\mu\Phi_\mu=\left(1-(1-\mu)R_\mu\right)\Phi_\mu\;.
\end{eqnarray}
The following lemma asserts that $\left(1-(1-\mu)R_\mu\right)$ is
invertible and under good control for $\mu\to 1$.
\begin{lem}\label{rmuop}
There exists $C<\infty$ such that
\begin{eqnarray}\label{rnorm}
\| R_\mu\|^{op}_2<C\left(1-\mu\right)^{-\frac{13}{16}}\;.
\end{eqnarray}
\end{lem}
{\bf Proof:} Let $\chi\in(\mathcal{N}\backslash\Phi)^\bot$ with
$\|\chi\|=1$. Set
\begin{equation}\label{defxi}\xi=P_{\mathcal{N}^\bot}\left(A\chi\right)\;.\end{equation}
i.e.
\begin{equation}\label{rallgemein}
R_\mu\chi=\left(D_1-E_\mu\right)^{-1}\xi\;.
\end{equation}
We observe that $\xi$ is orthogonal to $\mathcal{N}$. Set
\begin{equation}\label{cf2}
r_\mu=(1-\mu)^{-\frac{3}{8}}
\end{equation}
and let $\mathcal{B}_0(r_\mu)=\{x\in\mathbb{R}^3:\|x\|<r_\mu\}$.
Choose $\mu_B<1$ so that
$\mathcal{S}_A\subset\mathcal{B}_0(r_\mu)$ for all
$\mu\in[\mu_B,1]$. For large enough $r_\mu$ ($\mu_B$ close enough
to one) we have that the vectors $P_\mathcal{N}
(\mathds{1}_{\mathcal{B}(r_\mu)}\Phi^l)$, $1\leq l \leq n$ are
linearly independent. Hence there exists a
$\widetilde{\Phi}_\mu\in\mathcal{N}$ such that
\begin{equation}\label{projektion}
P_\mathcal{N}
(\mathds{1}_{\mathcal{B}(r_\mu)}\widetilde{\Phi}_\mu)=P_\mathcal{N}
(\mathds{1}_{\mathcal{B}(r_\mu)}\xi)\;.
\end{equation}
We define now the spacial cutoff parts
\begin{eqnarray}\label{defxi1}
\xi_{1,\mu}&:=&\mathds{1}_{\mathcal{B}(r_\mu)}\xi-\mathds{1}_{\mathcal{B}(r_\mu)}\widetilde{\Phi}_\mu
\\\label{defxi2}
\xi_{2,\mu}&:=&\xi-\xi_{1,\mu}\;,
\end{eqnarray}
which are orthogonal to $\mathcal{N}$. By Schwartz inequality
$$\|\xi_{1,\mu}\|_1\leq  \|\mathds{1}_{\mathcal{B}(r_\mu)}\|\|\xi_{1,\mu}\|\leq \left(\frac{4}{3}\pi r_\mu^3\right)^{\frac{1}{2}}\left(\|\xi\|+\|\mathds{1}_{\mathcal{B}(r_\mu)}\widetilde{\Phi}_\mu\|\right)\;.$$
Since $P_\mathcal{N} \widetilde{\Phi}_\mu=\widetilde{\Phi}_\mu$ we
have with (\ref{projektion})
\begin{eqnarray*}\|\mathds{1}_{\mathcal{B}(r_\mu)}\widetilde{\Phi}_\mu\|^2
&=&\langle\mathds{1}_{\mathcal{B}(r_\mu)}\widetilde{\Phi}_\mu,\widetilde{\Phi}_\mu\rangle
=\langle\mathds{1}_{\mathcal{B}(r_\mu)}\widetilde{\Phi}_\mu,P_\mathcal{N}
\widetilde{\Phi}_\mu\rangle
\\&=&\langle P_\mathcal{N} \left(\mathds{1}_{\mathcal{B}(r_\mu)}\widetilde{\Phi}_\mu\right),\widetilde{\Phi}_\mu\rangle
=\langle
\mathds{1}_{\mathcal{B}(r_\mu)}\xi,\widetilde{\Phi}_\mu\rangle
\\&=&\langle \mathds{1}_{\mathcal{B}(r_\mu)}\xi,\mathds{1}_{\mathcal{B}(r_\mu)}\widetilde{\Phi}_\mu\rangle
\leq\|\mathds{1}_{\mathcal{B}(r_\mu)}\xi\|\|\mathds{1}_{\mathcal{B}(r_\mu)}\widetilde{\Phi}_\mu\|\;,
%
%
\end{eqnarray*}
 hence
$$\|\mathds{1}_{\mathcal{B}(r_\mu)}\widetilde{\Phi}_\mu\|\leq\|\mathds{1}_{\mathcal{B}(r_\mu)}\xi\|\leq\|\xi\|\;,$$
and thus  since $\|\xi\|$ is finite there exists a constant
$C<\infty$ so that
\begin{equation}\label{l1normxi1}
\|\xi_{1,\mu}\|_1\leq Cr_\mu^{3/2}\;.
\end{equation}
For (\ref{rallgemein}) we obtain
$$R_\mu\chi=\left(D_1-E_\mu\right)^{-1}\xi_{1,\mu}+\left(D_1-E_\mu\right)^{-1}\xi_{2,\mu}$$
and we wish to show that for some $C<\infty$
\begin{eqnarray}\label{xi1est}
\|\left(D_{1}-E_\mu\right)^{-1}\xi_{k,\mu}\|<
C\left(1-\mu\right)^{-\frac{13}{16}}\;,\;\;\;k=1,2\;.
\end{eqnarray}
\underline{(\ref{xi1est}) $k=1$:} We introduce
$\Omega^j_0\in\mathcal{N}$ from Theorem \ref{gefthm}
(\ref{resultat2}). Since $\langle\xi_{1,\mu},\Omega^j_0\rangle=0$ we
have
\begin{eqnarray*}
\mathcal{F}_1\left(\xi^j_{1,\mu}\right)(\mathbf{k})&:=&(2\pi)^{-3/2}\langle\varphi^{j}(\mathbf{k},1,\mathbf{x}),\xi_{1,\mu}\rangle
\\&=&(2\pi)^{-3/2}\langle\varphi^{j}(\mathbf{k},1,\mathbf{x})-\Omega^j_{0}(\mathbf{k},1,\mathbf{x}),\xi_{1,\mu}\rangle\;.
\end{eqnarray*}
Thus
\begin{eqnarray*}
\left|\mathcal{F}_1\left(\xi^j_{1,\mu}\right)\right|\leq(2\pi)^{-3/2}\|\varphi^{j}(\mathbf{k},1,\cdot)-\Omega_{0}^j(\mathbf{k},1,\cdot)\|_\infty\;\|\xi_{1,\mu}\|_1\;.
\end{eqnarray*}
By using Theorem \ref{gefthm} and (\ref{l1normxi1}) we get
\begin{eqnarray}\label{xihut}
\left|\mathcal{F}_1\left(\xi^j_{1,\mu}\right)(\mathbf{k})\right|\leq
Cr_\mu^{3/2}\;.
\end{eqnarray}
Next note that with
$D_1\varphi^{j}(\mathbf{k},1,\cdot)=E_k\varphi^{j}(\mathbf{k},1,\cdot)=\sqrt{k^2+1}\varphi^{j}(\mathbf{k},1,\cdot)$
and $E_k\geq1>E_\mu$ and hence $E_k-E_\mu\geq 1-E_\mu>0$,
$E_k-E_\mu> E_k-1\geq0$
\begin{eqnarray*}
\|
\left(D_{1}-E_\mu\right)^{-1}\xi_{1,\mu}\|&=&\sum_j\|\frac{1}{E_k-E_\mu}\mathcal{F}_1\left(\xi^j_{1,\mu}\right)(\mathbf{k})\|
\\&\leq&\sum_j\|\frac{1}{E_k-E_\mu} \mathds{1}_{|k|<(1-E_\mu)^{1/2}}\mathcal{F}_1\left(\xi^j_{1,\mu}\right)(\mathbf{k})\|
\\&&+\sum_j\|\frac{1}{E_k-E_\mu} \mathds{1}_{1>|k|>(1-E_\mu)^{1/2}}\mathcal{F}_1\left(\xi^j_{1,\mu}\right)(\mathbf{k})\|
\\&&+\sum_j\|\frac{1}{E_k-E_\mu} \mathds{1}_{|k|>1}\mathcal{F}_1\left(\xi^j_{1,\mu}\right)(\mathbf{k})\|
\\&\leq&\sum_j\frac{1}{1-E_\mu}\|\mathds{1}_{|k|<(1-E_\mu)^{1/2}}\mathcal{F}_1\left(\xi^j_{1,\mu}\right)(\mathbf{k})\|
\\&&+\sum_j\| \mathds{1}_{1>|k|>(1-E_\mu)^{1/2}}\frac{1}{E_k-1}\mathcal{F}_1\left(\xi^j_{1,\mu}\right)(\mathbf{k})\|
\\&&+\sum_j\|\mathds{1}_{|k|>1}\frac{1}{E_k-1}
\mathcal{F}_1\left(\xi^j_{1,\mu}\right)(\mathbf{k})\|\;.
\end{eqnarray*}
By (\ref{xihut}) we obtain with appropriate constants $C_1$, $C_2$,
$C_3$
\begin{eqnarray*}
\| \left(D_{1}-E_\mu\right)^{-1}\xi_{1,\mu}\|
&\leq&\frac{C_1}{1-E_\mu}r_\mu^{3/2}(1-E_\mu)^{3/4}
\\&&+C_2r_\mu^{3/2}\left(\int\mathds{1}_{1>|k|>(1-E_\mu)^{1/2}}\left(E_k-1\right)^{-2}d^3k\right)^{1/2}
\\&&+C_3\;.
\end{eqnarray*}
Noting that $E_k-1\geq k^2/2$ for $|k|<1$, we obtain
\begin{eqnarray*}
%
\\&\leq&r_\mu^{3/2}\left(C_1(1-E_\mu)^{-1/4}+C_2
\left(4\pi\int_{(1-E_\mu)^{1/2}}^14k^{-2}dk\right)^{1/2}\right)
+C_3
\\&=&r_\mu^{3/2}\left(C_1(1-E_\mu)^{-1/4}+\frac{8\pi}{\sqrt{3}}C_2\left(\left(1-E_\mu\right)^{-1/2}-1\right)^{1/2}\right)+C_3\;.
\end{eqnarray*}
Hence there exists an appropriate constant $C<\infty$ such that
\begin{equation}\label{zeta1absch}\|\left(D_{1}-E_\mu\right)^{-1}\xi_{1,\mu}\|\leq
Cr_\mu^{3/2}(1-E_\mu)^{-1/4}\;.\end{equation} By
(\ref{diveproperly}) $1-E_\mu\geq \underline{C}(1-\mu)$. This and
(\ref{cf2})
yield (\ref{xi1est}) for $k=1$.\\

\noindent\underline{(\ref{xi1est})$k=2$:} By (\ref{defxi1}) and
(\ref{defxi2})
\begin{eqnarray*}
\|\xi_{2,\mu}\|=\|\xi-\xi_{1,\mu}\|&\leq&\|\xi-\mathds{1}_{\mathcal{B}(r_\mu)}\xi\|
+\|\mathds{1}_{\mathcal{B}(r_\mu)}\widetilde{\Phi}_\mu\|\;.
\end{eqnarray*}
Since  $\widetilde{\Phi}_\mu\in\mathcal{N}$ we have with
(\ref{projektion})
\begin{eqnarray*}|\langle \mathds{1}_{\mathcal{B}(r_\mu)}\xi,\widetilde{\Phi}_\mu\rangle|
&=&|\langle \mathds{1}_{\mathcal{B}(r_\mu)}\xi,P_\mathcal{N}
\widetilde{\Phi}_\mu\rangle|
=|\langle P_\mathcal{N}
\mathds{1}_{\mathcal{B}(r_\mu)}\xi,\widetilde{\Phi}_\mu\rangle|
\\&=&|\langle\mathds{1}_{\mathcal{B}(r_\mu)}\widetilde{\Phi}_\mu,\widetilde{\Phi}_\mu\rangle|
=\|\mathds{1}_{\mathcal{B}(r_\mu)}\widetilde{\Phi}_\mu\|^2\;.\end{eqnarray*}
On the other hand recalling that $\xi\bot\mathcal{N}$ we obtain by
Schwartz inequality
$$|\langle\mathds{1}_{\mathcal{B}(r_\mu)}\xi,\widetilde{\Phi}_\mu\rangle|=|\langle\xi-\mathds{1}_{\mathcal{B}(r_\mu)}\xi,\widetilde{\Phi}_\mu\rangle|\leq\|\xi-\mathds{1}_{\mathcal{B}(r_\mu)}\xi\|\;\;\|\widetilde{\Phi}_\mu\|\;,$$
hence
$$\|\mathds{1}_{\mathcal{B}(r_\mu)}\widetilde{\Phi}_\mu\|\leq C\|\xi-\mathds{1}_{\mathcal{B}(r_\mu)}\xi\|\frac{\|\widetilde{\Phi}_\mu\|}{\|\mathds{1}_{\mathcal{B}(r_\mu)}\widetilde{\Phi}_\mu\|}\;.$$
Clearly
$$\lim_{\mu\to1}\frac{\|\widetilde{\Phi}_\mu\|}{\|\mathds{1}_{\mathcal{B}(r_\mu)}\widetilde{\Phi}_\mu\|}=1\;,$$
thus for $\mu_B$ close enough to $1$ there exists a $C<\infty$ so
that
\begin{eqnarray}
\|\xi_{2,\mu}\|
&\leq&\|\xi-\mathds{1}_{\mathcal{B}(r_\mu)}\xi\|+\|\mathds{1}_{\mathcal{B}(r_\mu)}\widetilde{\Phi}_\mu\|%
\nonumber\\\label{einplus}&\leq&C\|\xi-\mathds{1}_{\mathcal{B}(r_\mu)}\xi\|\;.
\end{eqnarray}
By (\ref{defxi}) and the fact that $A$ has compact support we have
for $r_\mu$ large enough that $\xi$ is outside the ball
$\mathcal{B}(r_\mu)$ a multiple of $\Phi$. Hence
$\xi-\mathds{1}_{\mathcal{B}(r_\mu)}\xi$ is outside the ball
$\mathcal{B}(r_\mu)$ a multiple of $\Phi$. Since
$\Phi\in\mathcal{N}$ its decay properties are known from the Greens
function of $D_1-1$ (see e.g. \cite{klaus}, \cite{picklneu}), namely
$\mid\Phi\mid\leq Cx^{-2}$, we obtain
$$\|\xi_{2,\mu}\|\leq C\left(\int_{x>r_\mu} x^{-4}d^3x\right)^{\frac{1}{2}}\leq Cr_\mu^{-\frac{1}{2}}\;.$$
with appropriate constants $C<\infty$. It follows that
\begin{eqnarray}\label{xi2absch}
\| \left(D_{1}-E_\mu\right)^{-1}\xi_{2,\mu}\|
&=&\|
\frac{1}{E_k-E_\mu}\mathcal{F}_1\left(\xi_{2,\mu}\right)\|\leq\frac{1}{1-E_\mu}\|\xi_{2,\mu}\|
\nonumber\\&\leq&\mid 1-E_\mu\mid^{-1}C r_\mu^{-\frac{1}{2}}\;.
\end{eqnarray}
As before, (\ref{diveproperly}) and (\ref{cf2}) yield (\ref{xi1est})
for $k=2$.

 $\square$

\begin{cor}\label{thecorollary}
\hspace{1cm}
\begin{itemize}
\item[(i)] For $\mu_B$ close enough to $1$
\begin{equation}\label{sternchen}
\|(1-\mu)R_\mu\|_2^{op}\leq
(1-\mu)^{\frac{3}{16}}\;,\;\;\;\forall\;\mu\in[\mu_B,1)
\end{equation}
 and
$$\left(1-(1-\mu)R_\mu\right)^{-1}=\sum_{j=0}^\infty (1-\mu)^j R_\mu^j$$
exists as bounded operator on $(\mathcal{N}\backslash\Phi)^\bot$.

\item[(ii)] For $\mu\in[\mu_B,1)$ $\langle\Phi_\mu,\Phi\rangle\neq0$ and for
\begin{eqnarray}\label{zetadef}\zeta_\mu:=
\frac{\Phi_\mu}{\langle\Phi_\mu,\Phi\rangle}=\sum_{j=0}^\infty
(1-\mu)^jR_\mu^j\Phi
\end{eqnarray}
\begin{equation}\label{implieslem}
\lim_{\mu\to1}\|P_{\mathcal{N}^\bot}\zeta_{\mu}\|=\lim_{\mu\to1}\left\|\sum_{j=1}^\infty
(1-\mu)^jR_\mu^j\Phi\right\|=0\;
\end{equation}
holds.
\end{itemize}
\end{cor}

{\bf Proof}: (i) is immediate from the lemma. For (ii) observe
(\ref{rreihe}) and (i) to conclude that
$\langle\Phi_\mu,\Phi\rangle\neq0$. (\ref{implieslem}) follows
straightforwardly  from ( \ref{sternchen}).
 $\square$

 With this we establish now (i) of Lemma \ref{sequenz}: By
(\ref{implieslem}), (\ref{zetadef}) and observing that $\Phi_\mu$
and $\Phi$ are normalized
\begin{eqnarray*}
0=\lim_{\mu\to1}\frac{\|P_{\mathcal{N}^\bot}\Phi_\mu\|}{|\langle\Phi_\mu,\Phi\rangle|}
\geq\lim_{\mu\to1}\|P_{\mathcal{N}^\bot}\Phi_\mu\|
\end{eqnarray*}
which implies (i) of Lemma \ref{sequenz}.

Next we prove (ii) of Lemma \ref{sequenz}. Let $n_\mu:=\dim \mathcal{N}_\mu$. Let $\{\Phi^l,l=1,\ldots,n\}$ be a
basis of $\mathcal{N}$, $\{\Phi^l_\mu,l=1,\ldots,n_\mu\}$ be a basis of $\mathcal{N}_\mu$.

We first show by contradiction that for $\mu_B$ close enough to $1$ $n\geq n_\mu$ for all $\mu\in[\mu_B,1)$.
Assume that for any $0<\mu_B<1$ there exists a $\mu\in[\mu_B,1)$ such that $n<n_\mu$. Then the $n$-dimensional
vectors $\{v_j, j=1,\ldots,n_\mu\}$ defined by their coordinates $v_j^l:=\langle\Phi^j_\mu,\Phi^l\rangle$ are
linearly dependent, i.e. there exist nontrivial $\alpha_j$ such that $\sum_{j=1}^{n_\mu}\alpha_j v_j=0$. Then
$\Phi_\mu:=\sum_{j=1}^{n_\mu}\alpha_j \Phi_\mu^j\neq 0$ satisfies $P_\mathcal{N}\Phi_\mu=0$. Hence for any
$0<\mu_B<1$ there exists a $\mu\in[\mu_B,1)$ and a $\Phi_\mu\in\mathcal{N}_\mu$ such that
$P_\mathcal{N}\Phi_\mu=0$. This contradicts part (i) of Lemma \ref{sequenz}. Hence $n\geq n_\mu$.

Next we show that for $\mu_B$ close enough to $1$ $n\leq n_\mu$ for all $\mu\in[\mu_B,1)$. Again assume that for
any $0<\mu_B<1$ there exists a $\mu\in[\mu_B,1)$ such that $n>n_\mu$. Analogously as above we obtain that for
any $0<\mu_B<1$ there exists a $\mu\in[\mu_B,1)$ and a $\Phi\in\mathcal{N}$ such that
$P_{\mathcal{N}_\mu}\Phi=0$.

But for any $\Delta E\in\mathbb{R}$ we have
\begin{eqnarray*}
\|(D_\mu-1+\Delta E)\Phi\|^2&=&\|(D_0+(\mu-1)A-1+\Delta E)\Phi\|^2\\&=&
\langle(\Delta E-(1-\mu)A)^2\Phi,\Phi\rangle
\\&=&(\Delta E)^2+\langle(-2\Delta E
(1-\mu)A+(1-\mu)^2A^2)\Phi,\Phi\rangle\;.
\end{eqnarray*}
Choosing $\Delta E=(1-\mu)\|A\|_\infty<1$  for $\mu_B$ close enough to $1$, it follows that
\begin{equation}\label{contra}\|(D_\mu-1+\Delta E)\Phi\|^2<(\Delta E)^2\;.\end{equation} On the other hand
$P_{\mathcal{N}_\mu}\Phi=0$ implies by virtue of Condition \ref{cond} (ii) that $\Phi$ lies in the absolutely
continuous spectrum. Writing $P^+_\mu$ and $P^-_\mu$ for the spectral projections onto the positive and negative
absolutely continuous spectral subspaces of $D_\mu$ and using that $\Delta E<1$ we obtain
\begin{eqnarray*}\|(D_\mu-1+\Delta E)\Phi\|^2&=&\|(D_\mu-1+\Delta
E)P^+_\mu\Phi\|^2+\|(D_\mu-1+\Delta E)P^-_\mu\Phi\|^2
\\&\geq&\|(1-1+\Delta E)P^+_\mu\Phi\|^2+\|(-1-1+\Delta
E)P^-_\mu\Phi\|^2\\&\geq&(\Delta E)^2\|P^+_\mu\Phi\|^2+\|P^-_\mu\Phi\|^2
\\&\geq&(\Delta
E)^2\|P^+_\mu\Phi\|^2+(\Delta E)^2\|P^-_\mu\Phi\|^2
=(\Delta E)^2\;.\end{eqnarray*} This contradicts to (\ref{contra}) and hence $n=n_\mu$.

$\square$

\subsection{$\mu$-Derivative of the Bound States}\label{51}
We shall construct now for each element $\Phi\in\mathcal{N}$ a
sequence of elements $(\Phi_\mu)_\mu$ in $(\mathcal{N}_\mu)_\mu$
which is ``good'' in several respects:
\begin{lem}\label{goodseq}
For $\mu_B$ close enough to $1$ holds: For each $\Phi\in
\mathcal{N},\,\,\|\Phi\|=1, $ exists a unique sequence
$(\zeta_\mu)_{\mu\in[\mu_B,1)}, \,\, \zeta_\mu\in\mathcal{N}_\mu, $
for which
\begin{itemize}
\item[(i)] $\langle\zeta_\mu, \Phi\rangle=1$,  \item[(ii)] $\zeta_\mu\in (\mathcal{N}\backslash\Phi)^\bot$,
where $(\mathcal{N}\backslash\Phi)^\bot$ is the orthogonal complement of $\mathcal{N}\backslash\Phi$.
\end{itemize}
We call the sequence $\Phi_\mu:=\frac{\zeta_\mu}{\|\zeta_\mu\|}$ a
good sequence corresponding to $\Phi$.
\end{lem}
{\bf Proof:} Choose for each $\mu\in[\mu_B,1)$ an orthonormal basis
$\{\zeta_\mu^k,\, k=1,\ldots,n\}$ of $\mathcal{N}_\mu$. Then by
(\ref{wahr})
\begin{equation}\label{5.1}
\lim_{\mu\to 1}\|P_\mathcal{N}\zeta_\mu^k\|=1, \,\,k=1,\ldots,n.
\end{equation}
Decompose the vector we are looking for as
\begin{equation}\label{5.2}
\zeta_\mu = \sum_{k=1}^n \alpha_\mu^k\zeta_\mu^k\,.
\end{equation}
Introduce an orthonormal basis $\{\Phi^1=\Phi, \Phi^k,\,
k=2,\ldots,n\}$ of $\mathcal{N}$. Then (i) and (ii) of the lemma
read
\begin{eqnarray*}
\sum_{k=1}^n \alpha_\mu^k\langle \zeta_\mu^k\,\Phi\rangle &=& 1\\
\sum_{k=1}^n \alpha_\mu^k \langle \zeta_\mu^k\,\Phi^l\rangle&=&0
\,\,, l=2,\ldots,n.
\end{eqnarray*}
This is a linear system of $n$ equations for the vector $
(\alpha_\mu)$ with matrix $M_\mu$ given by
$$M_\mu(i,j)=\langle\zeta_\mu^i,\Phi^j\rangle.$$ We note that the
column vectors of $M$ are the coordinates of the vectors
$P_\mathcal{N}\zeta_\mu^k$ in the orthonormal basis $\{\Phi^1=\Phi,
\Phi^k,\, k=2,\ldots,n\}$. They are linearly independent: Suppose
that were not so, then there exists a sequence $\mu_p$ converging to
$1$ so that $$\sum_{k=1}^n
\lambda_{\mu_p}^kP_\mathcal{N}\Phi_{\mu_p}^k=0\,$$ with
$(\lambda^1_{\mu_p},...,\lambda^n_{\mu_p})\neq 0,$ i.e. there exists
a sequence of normalized vectors
$\widetilde{\Phi}_{\mu_p}:=\frac{\sum_{k=1}^n
\lambda_{\mu_p}^k\Phi_{\mu_p}^k}{\|\sum_{k=1}^n
\lambda_{\mu_p}^k\Phi_{\mu_p}^k\|} \in\mathcal{N}_\mu$ for which
$\|P_\mathcal{N}\widetilde{\Phi}_{\mu_p}\|=0,$ contradicting
(\ref{5.1}). Thus $M_\mu$ is invertible and we find
\begin{equation}\label{5.3}
\zeta_\mu=\sum_{k=1}^n M_{\mu}^{-1}(k,1)\Phi_{\mu}^k.
\end{equation}
Thus (i) and (ii) are satisfied.

\begin{lem}\label{lemmacbder}($\mu$-Derivative of the Bound States)
Let $\Phi\in\mathcal{N}$ normalized and let $\Phi_\mu$,
$\mu\in[\mu_B,1]$ be a good sequence corresponding to $\Phi$.
Then\footnote{This estimate is not optimal, but sufficient for
what is needed later. It seems reasonable to conjecture that the
correct exponent is $-\frac{1}{2}$.}
\begin{equation}\label{lemcbderb}\|\partial_\mu \Phi_\mu\|\leq C
(1-\mu)^{-\frac{13}{16}}\;,\;\;\;\mu\in[\mu_B,1]
\end{equation}
with some constant $C<\infty$. Since $\mathcal{N}$ is finite
dimensional the constant can be chosen uniformly on $\mathcal{N}$.
\end{lem}
{\bf Proof:} Let $\Phi\in\mathcal{N}$ and let $\Phi_\mu$,
$\mu\in[\mu_B,1]$ be a good sequence corresponding to $\Phi$, i.e.
$\|\Phi\|=1$ and $\Phi_\mu\in(\mathcal{N}\backslash\Phi)^\bot$. At
the beginning of the proof of Lemma \ref{sequenz} we said that we
shall adjust the proof for later reference. We shall now use some
definitions and results of that proof here. The important
observation is that $\Phi$ and its corresponding good sequence are
in exact correspondence to the a priori given $\Phi_\mu$ in Lemma
\ref{sequenz} and the a posteriori chosen $\Phi$ as defined in the
beginning of the proof of Lemma \ref{sequenz}. With the good
sequence we invert now the situation: $\Phi$ is given and $\Phi_\mu$
is chosen. We shall use the formula (\ref{zetadef}) to differentiate
$\Phi_\mu$ (given by $\zeta_\mu$ (\ref{zetadef})). Formally
\begin{eqnarray*}
\partial_\mu\zeta_{\mu}&=&\partial_\mu\sum_{j=0}^\infty \left((\mu-1)R_{\mu}\right)^j\Phi
\\&=&(\partial_\mu (\mu-1)R_{\mu})\sum_{j=1}^\infty j\left((\mu-1)R_{\mu}\right)^{j-1}\Phi
\\&=&R_{\mu}\sum_{j=1}^\infty
j\left((\mu-1)R_{\mu}\right)^{j-1}\Phi
\\&&+(\mu-1)(\partial_\mu R_{\mu})\sum_{j=1}^\infty
j\left((\mu-1)R_{\mu}\right)^{j-1}\Phi\;.
\end{eqnarray*}
Hence by Corollary \ref{thecorollary} we obtain
rigorously\begin{eqnarray}\label{dszeta}
\|\partial_\mu\zeta_{\mu}\|&\leq&C\left(\|
R_{\mu}\|_2^{op}+\|(\mu-1)\partial_\mu R_{\mu}\|_2^{op}\right)\;.
\end{eqnarray}
For the second term observe that  formally
\begin{eqnarray*}
\partial_\mu R_\mu \chi&=&\partial_\mu\left(D_{1}-E_\mu\right)^{-1}\left(A\chi-\langle A\chi,\Phi\rangle\Phi\right)
\\&=&\left(\partial_\mu E_\mu\right)\left(D_{1}-E_\mu\right)^{-2} \left(A\chi-\langle A\chi,\Phi\rangle\Phi\right)
\\&=&\left(\partial_\mu E_\mu\right)\left(D_{1}-E_\mu\right)^{-1} R_\mu\chi\;,
\end{eqnarray*}
which can  be justified using for example the spectral
decomposition of $D_1$. From this we get
\begin{equation*} \|\partial_\mu R_\mu\|_2^{op}\leq C(1-\mu)^{-1}\| R_\mu\|_2^{op}
\end{equation*}
hence
\begin{eqnarray}\label{dsrnorm}
\|\partial_\mu\zeta_{\mu}\|&\leq&C\| R_\mu\|_2^{op}\;.
\end{eqnarray}

Finally, since by (\ref{zetadef})
$$\Phi_\mu=\frac{\zeta_{\mu}}{\| \zeta_{\mu} \|}\;,$$
$$\partial_\mu\Phi_\mu=\frac{\partial_\mu\zeta_{\mu}}{\| \zeta_{\mu}
\|}-\frac{\zeta_{\mu}}{\| \zeta_{\mu} \|^2}\partial_\mu \| \zeta_{\mu} \|\;.$$ Furthermore we have that $\|
\zeta_{\mu}\|\geq 1$ for all $\mu\in[\mu_B,1[$ and by triangle inequality
$\partial_\mu\|\zeta_{\mu}\|\leq\|\partial_\mu\zeta_{\mu}\|$, therefore

$$\|\partial_\mu\Phi_\mu\|\leq C\| R_\mu\|_2^{op}$$
and (\ref{lemcbderb}) follows with Lemma \ref{rmuop}.
\begin{flushright}$\Box$\end{flushright}


\section{Proof of Theorem \ref{onecrit}}\label{proofmainlemma}

We now study the true time evolution $U^\varepsilon(0,s)$ as given
by (\ref{U}). To prove Theorem \ref{onecrit} we have to control
the time propagation of $\psi_{s}^\varepsilon$. This propagation
is naturally qualitatively different for $s<0$ (``adiabatic bound
state evolution'') and $s>0$(``scattering''). Hence we control the
propagation for $s<0$ and $s>0$ separately.

\subsection{Control of $\psi_{s}^\varepsilon$ for $s_0\leq s\leq 0$}\label{secco}
Usually adiabatic theory assumes a spectral gap. Here we are in a
situation where the eigenvalue $E_{\mu(s)}$ will close the gap to
the upper continuum. We need to control the adiabatic change of the
bound states without a gap condition.

\begin{lem}\label{corvor0} (Adiabatic Lemma without a gap)\\
Let $s<0$ be such that a bound state  $\widetilde{\Phi}_{\mu(s)}$ of $D_{\mu(s)}$ with energy $E_{\mu(s)}>-1$
exists. Then

\begin{eqnarray}\label{corvoreq1}
\lim_{\varepsilon\to0}\| P_\mathcal{N} U^\varepsilon(0,s)\widetilde{\Phi}_{\mu(s)}\|&=&1\;.
\end{eqnarray}

\end{lem}

\noindent\textbf{Proof:} Let $s<0$, $\widetilde{\Phi}_{\mu(s)}$ be a
bound state. By the adiabatic Theorem \cite{teufel} we have that for
any $s_0<1$
\begin{equation}\label{adiatheo}\lim_{\varepsilon\to0}\|
P_{\mathcal{N}_{\mu(s_0)}}U^\varepsilon(s_0,s)\widetilde{\Phi}_{\mu(s)}\|=1\;.
\end{equation}
Setting
$\widetilde{\Phi}^\varepsilon_{\mu(s_0)}:=P_{\mathcal{N}_{\mu(s_0)}}U^\varepsilon(s_0,s)\widetilde{\Phi}_{\mu(s)}$
we note that (\ref{adiatheo}) is equivalent to
\begin{equation}\label{adiatheo1}\lim_{\varepsilon\to0}
\|U^\varepsilon(s_0,s)\widetilde{\Phi}_{\mu(s)}-\widetilde{\Phi}^\varepsilon_{\mu(s_0)}\|=0\;.
\end{equation}
  Let $\{\Phi^l\}$ be an orthonormal basis of $\mathcal{N}$.
Consider the corresponding good sequences $\Phi_\mu^l$ which by
definition in Lemma \ref{goodseq} and by (\ref{wahr}) satisfy
$\lim_{\mu\to 1}\langle\Phi_\mu^l,\Phi^l\rangle=1,
\,\,\,\langle\Phi_\mu^l,\Phi^k\rangle=0,\,\,k\ne l.$ It thus follows
 that for every $\mu\in[\mu_B,1),$
$\{\Phi_\mu^k,\,k=1,\dots,n\} $ forms a basis in $\mathcal{N}_\mu.$
Let now  $s_0$ be such that $\mu(s_0)\ge \mu_B$.  We decompose the
bound state
$$\widetilde{\Phi}^\varepsilon_{\mu(s_0)}
= \sum_{l=1}^n\alpha^\varepsilon_{s_0,l}\Phi_{\mu(s_0)}^l\,.$$ For
this basis
\begin{equation}\label{alphasum}
\lim_{s_0\to 0}
|\langle\Phi_{\mu(s_0)}^l,\Phi_{\mu(s_0)}^k\rangle|=\delta_{k,l}\,\,,
\end{equation}
since for $l\neq k$
$$\lim_{s_0\to 0}
|\langle\Phi_{\mu(s_0)}^l,\Phi_{\mu(s_0)}^k\rangle|=\lim_{s_0\to
0}
|\langle\Phi_{\mu(s_0)}^l-\Phi^l,\Phi_{\mu(s_0)}^k\rangle|\le\lim_{s_0\to
0} \|\Phi_{\mu(s_0)}^l-\Phi^l\|\|\Phi_{\mu(s_0)}^k\|$$ and
$$\lim_{s_0\to0}\|\Phi_{\mu(s_0)}^l-\Phi^l\|^2=1+1-\lim_{s_0\to0}\langle\Phi_{\mu(s_0)}^l,\Phi^l\rangle-\lim_{s_0\to0}\langle\Phi^l,\Phi_{\mu(s_0)}^l\rangle=0.$$
 From (\ref{alphasum})  we can conclude that
\begin{equation}\label{alphasum1}
\overline{\lim_{s_0\to 0}}\sum_{k=1}^n
|\alpha^\varepsilon_{s_0,k}|^2\leq 1.
\end{equation}
Now use the coordinates $\alpha^\varepsilon_{s_0,l}$ to define an
approximate time evolution
\begin{equation}\label{defpsi1k}
\Phi_{\mu(s),s_0}^\varepsilon:=
\exp\left(-\frac{i}{\varepsilon}\int_{s_0}^{s}E_{\mu(v)}dv\right)\sum_{l=1}^n\alpha^\varepsilon_{s_0,l}\Phi_{\mu(s)}^l\;.
\end{equation}
We note that $\Phi_{\mu(0)}^l=\Phi^l$ and thus
\begin{equation}\label{defpsi1kneu}
\Phi_{\mu(0),s_0}^\varepsilon=
\exp\left(-\frac{i}{\varepsilon}\int_{s_0}^{0}E_{\mu(v)}dv\right)\sum_{l=1}^n\alpha^\varepsilon_{s_0,l}\Phi^l\;.
\end{equation}
We compare the approximate time evolution with the true one
\begin{eqnarray*}
&&\hspace{-0.5cm}\Phi_{\mu(0),s_0}^\varepsilon-U^\varepsilon(0,s_0)\widetilde{\Phi}_{\mu(s_0)}=
\int_{s_0}^0\partial_u\left(U^\varepsilon(0,u)\Phi_{\mu(u),s_0}^\varepsilon\right)du
\\&&=-i
\int_{s_0}^0U^\varepsilon(s,u)\left(\frac{D_{u}}{\varepsilon}-i\partial_{u}\right)
\exp\left(-\frac{i}{\varepsilon}\int_{s_0}^uE_{\mu(v)}dv\right)\sum_{l=1}^n\alpha^\varepsilon_{s_0,l}\Phi_{\mu(u)}^l
du
\\&&=-\frac{i}{\varepsilon}
\int_{s_0}^0U^\varepsilon(s,u)\left(D_{u}-E_{\mu(u)}\right)
\exp\left(-\frac{i}{\varepsilon}\int_{s_0}^uE_{\mu(v)}dv\right)\sum_{l=1}^n\alpha^\varepsilon_{s_0,l}\Phi_{\mu(u)}^l
du
\\&&\;\;\;+i
\int_{s_0}^0U^\varepsilon(s,u)\exp\left(-\frac{i}{\varepsilon}\int_{s_0}^uE_{\mu(v)}dv\right)
\sum_{l=1}^n\alpha^\varepsilon_{s_0,l}\partial_u\Phi_{\mu(u)}^l
du\;.
\end{eqnarray*}
Since $(D_u-E_{\mu(u)})\Phi_{\mu(u)}=0$
\begin{eqnarray*}
\Phi_{\mu(0),s_0}^\varepsilon&-&U^\varepsilon(0,s_0)\widetilde{\Phi}_{\mu(s_0)}\\
&=&i
\int_{s_0}^0U^\varepsilon(s,u)\exp\left(-\frac{i}{\varepsilon}
\int_{s_0}^uE_{\mu(v)}dv\right)\sum_{l=1}^n\alpha^\varepsilon_{s_0,l}\partial_u\Phi_{\mu(u)}^ldu\;.
\end{eqnarray*}
Hence by unitarity of $U^\varepsilon$, Lemma \ref{lemmacbder} and Condition \ref{cond} (ii)
\begin{eqnarray}\label{uebersehen}
\|\Phi_{\mu(0),s_0}^\varepsilon-U^\varepsilon(0,s_0)\widetilde{\Phi}_{\mu(s_0)}\|&\leq&\sum_{l=1}^n|\alpha^\varepsilon_{s_0,l}|
\int_{s_0}^0\|\partial_u\Phi_{\mu(u)}^l\| du\nonumber
\\&\leq&
C\int_{s_0}^0u^{-\frac{13}{16}} du= \frac{13}{16}Cs_0^{\frac{3}{16}}\;,
\end{eqnarray}
where we concluded from (\ref{alphasum1})  that
$\sum_{l=1}^n|\alpha^\varepsilon_{s_0,l}|$ is bounded for $s_0$
close enough to one. Furthermore we obtain from (\ref{uebersehen})
that
\begin{equation}\nonumber
\lim_{s_0\to
0}\lim_{\varepsilon\to0}|\|\Phi_{\mu(0),s_0}^\varepsilon\|-\|U^\varepsilon(0,s_0)\widetilde{\Phi}_{\mu(s_0)}\||=\lim_{s_0\to
0}\lim_{\varepsilon\to0}|\|\Phi_{\mu(0),s_0}^\varepsilon\|-1|=0\,,
\end{equation} so that
\begin{equation}\label{brauchtman}
\lim_{s_0\to
0}\lim_{\varepsilon\to0}\|\Phi_{\mu(0),s_0}^\varepsilon\|=1\,.
\end{equation}

 Since $\Phi_{\mu(0),s_0}^\varepsilon\in\mathcal{N}$
\begin{eqnarray*}
\left|\| P_\mathcal{N} U^\varepsilon(0,s)\widetilde{\Phi}_{\mu(s)}\|-\|\Phi_{\mu(0),s_0}^\varepsilon\|\right|
&\leq&\| P_\mathcal{N} U^\varepsilon(0,s)\widetilde{\Phi}_{\mu(s)}-P_\mathcal{N} \Phi_{\mu(0),s_0}^\varepsilon\|
\\&\leq&\|U^\varepsilon(0,s)\widetilde{\Phi}_{\mu(s)}-U^\varepsilon(0,s_0)\widetilde{\Phi}_{\mu(s_0)}\|
\\&&+\|U^\varepsilon(0,s_0)\widetilde{\Phi}_{\mu(s_0)}-\Phi_{\mu(0),s_0}^\varepsilon\|
\\&=&\|U^\varepsilon(s_0,s)\widetilde{\Phi}_{\mu(s)}-\widetilde{\Phi}_{\mu(s_0)}\|
\\&&+\|U^\varepsilon(0,s_0)\widetilde{\Phi}_{\mu(s_0)}-\Phi_{\mu(0),s_0}^\varepsilon\|\;.
\end{eqnarray*}
It follows with   (\ref{adiatheo1}), (\ref{uebersehen}) and
(\ref{brauchtman}), that
\begin{eqnarray*}
\lim_{\varepsilon\to0}\| P_\mathcal{N} U^\varepsilon(0,s)\widetilde{\Phi}_{\mu(s)}\|&=&
\lim_{s_0\to0}\lim_{\varepsilon\to0}\| P_\mathcal{N} U^\varepsilon(0,s)\widetilde{\Phi}_{\mu(s)}\|
\\&=&\lim_{s_0\to0}\lim_{\varepsilon\to0}\|
\Phi_{\mu(0),s_0}^\varepsilon\|=1\;.
%
\end{eqnarray*}

\begin{flushright}$\Box$\end{flushright}

\subsection{Propagation Estimates for the Time Dependent Case: ``Short''
Times}\label{secpe1}

We shall introduce a time $\sigma>0$ which is a time of order one.
For example the time at which the switching factor $\mu(s)$ is
half way between $1$ and its maximum. Our estimates will be valid
until this time. It is in fact the crucial time after which the
critical bound state has already left the range of the potential.
We shall in the next section consider ``long'' times, i.e. the
times bigger than $\sigma$.
 We shall now
consider the auxiliary time evolution (\ref{V}) on the macroscopic
time scale $s=t\varepsilon$. That evolution will be denoted by
 $V^\varepsilon_{\mu(v)}(s,0)$ where $v$ is fixed! It is defined
by
\begin{equation}\label{Veps}
i\partial_s V_{\mu(v)}^\varepsilon(s,0)=
\frac{1}{\varepsilon}D_{\mu(v)} V_{\mu(v)}^\varepsilon(s,0)\,.
\end{equation}

We first reformulate our Corollary \ref{propestalt2} for
$V^\varepsilon_{\mu(v)}$ be given by (\ref{Veps}). Instead of
$\mu\in (1,1+\delta]$ we have now $v\in(0,\sigma]$. For the chosen
  $\sigma$  we can replace in view of
(\ref{diveproperly}) the factor $\mu(v)-1$ corresponding to
$\mu-1$ in Corollary \ref{propestalt2} by $v$ at little extra
costs.  We formulate first this adjustment as

\begin{cor}\label{propest}(Propagation Estimate - stationary case)\\
Let  $\mathcal{S}\subset\mathbb{R}^3$ be compact. There exists $\sigma>0$ such that for all
$\widetilde{m}\in\mathbb{N}$ and for all $0<\xi<1$ exist constants $C_{\xi,\widetilde{m}}$ and $C_\xi$ such that
for all $v\in(0,\sigma]$, all $u>\varepsilon(\mu(v)-1)^{-\frac{3}{2(1-\xi)}}$ and all $\chi\in L^2$ with ${\rm
supp}\chi\subset\mathcal{S}$ the following holds
\begin{itemize}
\item[(i)] for $\underline{\kappa}=\varepsilon^{\frac{1}{2}(1-\xi)} u^{-\frac{1}{2}(1-\xi)}$ and for all $s\geq
u$
\begin{eqnarray*}
\|\mathds{1}_\mathcal{S} V^\varepsilon_{\mu(v)}(s,0)\rho_{\underline{\kappa},\mu(v)}\chi\|
&\leq&C_{\xi,\widetilde{m}}(\|D_{\mu(v)}\chi\|)\varepsilon^{\widetilde{m}}s^{-\widetilde{m}}\;,
\end{eqnarray*}

\item[(ii)] for $\underline{\kappa}=\varepsilon^{\frac{1}{2}(1-\xi)}u^{-\frac{1}{2}(1-\xi)}$
$$\|(1-\rho_{\underline{\kappa},\mu(v)})\chi\|\leq C_{\xi}\varepsilon^{\frac{3}{4}(1-\xi)}u^{-\frac{3}{4}(1-\xi)}v^{-1/4}\|\chi\|\;,$$

\item[(iii)]
\begin{eqnarray*}
\|\mathds{1}_\mathcal{S} V^\varepsilon_{\mu(v)}(u,0)\chi\|
&\leq&C_\xi(\|\chi\|+\|D_{\mu(v)}\chi\|)v^{-1/2}\varepsilon^{\frac{3}{2}(1-\xi)}u^{-\frac{3}{2}(1-\xi)}\;.
\end{eqnarray*}
\end{itemize}
\end{cor}
We shall use that to control the time evolution of a wavefunction under the influence of the time dependent
Dirac operator.

\begin{lem}\label{timeprop}(Propagation Estimates - Time Dependent Case: ``Short'' times)\\
Let $U^\varepsilon(s,u)$ be given by (\ref{U}). Let $\chi\in L^2$ be normalized with ${\rm
supp}\chi\subset\mathcal{S}$ and finite energy, i.e. $\|D_0\chi\|<\infty$. Let $\sigma>0$ be as in Corollary
\ref{propest} and such that $\partial_s \mu(s)\geq \underline{C}>0$ on $(0,\sigma]$. For all $0<\xi<1/3$ exist
$C_\xi$ such that for all $s\in (0,\sigma]$
\begin{equation}\label{onecriteqb}
\|\mathds{1}_\mathcal{S}U^\varepsilon(s,0)\chi\|\leq
C_\xi\left(\varepsilon^{\frac{1}{2}-\frac{3}{2}\xi}s^{-\frac{3}{2}}\right)\;.
\end{equation}
\end{lem}
\begin{rem}
This estimate gives the decay time of the critical bound state. It is of the order of $\varepsilon^{1/3}$, i.e.
$\varepsilon^{-2/3}$ on the microscopic time scale. One should compare this with the decay of an $L^2$-function
in a non-critical situation which is of order one on the microscopic time scale.
\end{rem}
{\bf Proof:} Using that $\chi$ is normalized the Lemma follows trivially for $s\leq\varepsilon^{1/3-\xi}$ by
choosing $C_\xi>1$. Let $s>\varepsilon^{1/3-\xi}$ and
\begin{equation}\label{psiii}
\psi^{\varepsilon}_s:=U^\varepsilon(s,0)\chi.
\end{equation}
Now  $V_{\mu(s)}^\varepsilon$  is controllable with help of Corollary \ref{propest}. We shall ``replace'' the
propagator $U^\varepsilon$ by $V_{\mu(v)}^\varepsilon$. Then
$$\partial_s U^\varepsilon(s,0)=\varepsilon^{-1}D_sU^\varepsilon(s,0)\;\;\;\partial_s V_{\mu(v)}^\varepsilon(s,0)=\varepsilon^{-1}D_{\mu(v)}V_{\mu(v)}^\varepsilon(s,0)\;.$$
and
\begin{eqnarray}
U^\varepsilon(s,0)-V_{\mu(v)}^\varepsilon(s,0)&=&\int_{0}^s
\partial_u\left(V_{\mu(v)}^\varepsilon(s,u)U^\varepsilon(u,0)\right)du
\\\nonumber&=&-\frac{i}{\varepsilon}\int_{0}^sV_{\mu(v)}^\varepsilon(s,u)\left(D_{\mu(v)}-D_{\mu(u)}\right)U^\varepsilon(u,0)du
\\\nonumber\label{cook}&=&-\frac{i}{\varepsilon}\int_{0}^sV_{\mu(v)}^\varepsilon(s,u)\left(\mu(v)-\mu(u)\right)A(x)U^\varepsilon(u,0)du\;.
\end{eqnarray}
Hence
\begin{eqnarray}\label{cookwave}
\psi^{\varepsilon}_s&=&U^\varepsilon(s,0)\chi\nonumber\\&=&
%
V_{\mu(v)}^\varepsilon(s,0)\chi
+\frac{i}{\varepsilon}\int_{0}^s\left(\mu(u)-\mu(v)\right)V_{\mu(v)}^\varepsilon(s,u)A(x)\psi_u^\varepsilon
du\;.
\end{eqnarray}

We shall now choose a ``good'' $v$. The good choice is $v=s$. We
shall explain why: The ``error'' coming from
$\left(\mu(s)-\mu(u)\right)A(x)\psi_u^\varepsilon$ for $u$ close to
$s$ is very small. The ``error'' coming from earlier times is large
in $L^2$, but the propagation time $s-u$ is also large and hence
most of the wavefunction will have left the region $\mathcal{S}$
(c.f. Corollary \ref{propest})). So our strategy is not to show that
the ``error'' is small in $L^2$ (which would not work) but to show
that the ``error'' which is not small in $L^2$ leaves the region
$\mathcal{S}$ and what is left of the error in the relevant region
is small and thus in fact deserves to be called an error. The
estimates in Corollary \ref{propest} are only valid from a small
time on. This is an inheritance of the singular behavior of the
generalized eigenfunctions and must be taken into account. This will
lead to a slight complication which makes another splitting
necessary.

In detail: choosing $u=s$ we obtain, splitting the time according
to the idea above introducing another cutoff $\widetilde{\sigma}$
which will be specified below (and which takes care of the
applicability of the Corollary \ref{propest})
\begin{eqnarray*}
\psi^{\varepsilon}_s&=&V_{\mu(s)}^\varepsilon(s,0)\chi
+\frac{i}{\varepsilon}\int_{s-\widetilde{\sigma}}^s\left(\mu(u)-\mu(s)
\right)V_{\mu(s)}^\varepsilon(s,u)A(x)\psi_u^\varepsilon
du
\\&&+\frac{i}{\varepsilon}\int_{0}^{s-\widetilde{\sigma}}\left(\mu(u)-
\mu(s)\right)V_{\mu(s)}^\varepsilon(s,u)A(x)\psi_u^\varepsilon du\;.
\end{eqnarray*}
Hence
\begin{eqnarray}\label{props}
\nonumber\|\mathds{1}_\mathcal{S}\psi^{\varepsilon}_s\|&\leq&\|\mathds{1}_\mathcal{S}V_{\mu(s)}^\varepsilon(s,0)\chi\|
+\frac{1}{\varepsilon}\int_{s-\widetilde{\sigma}}^s\left(\mu(s)-\mu(u)\right)\|A(x)\psi_u^\varepsilon \|du
\\&&+\frac{1}{\varepsilon}\int_{0}^{s-\widetilde{\sigma}}\left(\mu(s)-\mu(u)\right)\|\mathds{1}_\mathcal{S}V_{\mu(s)}^\varepsilon(s,u)A(x)\psi_u^\varepsilon\|
du\;.
\end{eqnarray}
In view of (\ref{diveproperly}) the second summand is bounded
 by
\begin{eqnarray}\label{boundonsecond}
\frac{C}{\varepsilon}\int_{s-\widetilde{\sigma}}^s\left(s-u\right)\|A(x)\|_\infty\|\mathds{1}_{\mathcal{S}_A}\psi_u^\varepsilon
\|du
\leq \frac{C}{\varepsilon} \widetilde{\sigma}^2\;.
\end{eqnarray}
For the other terms we want to use Corollary \ref{propest} (iii). Therefore we have to control
$\|D_sA\psi^\varepsilon_u\|$, which we will do next. We have that (the differential symbol $\partial_u$ stands
also for $\frac{\rm d}{\rm d u}$)
\begin{eqnarray*}
&&\hspace{-0.5cm}\left|\partial_s\langle\psi^{\varepsilon}_s,D_{\mu(s)},\psi^{\varepsilon}_s\rangle\right|
\\&&=\left|\langle\psi^{\varepsilon}_s,(\partial_sD_{\mu(s)}),\psi^{\varepsilon}_s\rangle
+\langle(\partial_s\psi^{\varepsilon}_s),D_{\mu(s)},\psi^{\varepsilon}_s\rangle
+\langle\psi^{\varepsilon}_s,D_{\mu(s)},\partial_s\psi^{\varepsilon}_s\rangle\right|
\\&&=\left|\langle\psi^{\varepsilon}_s,A(\partial_s\mu(s)),\psi^{\varepsilon}_s\rangle
+\langle\frac{i}{\varepsilon}D_{\mu(s)}\psi^{\varepsilon}_s,D_{\mu(s)},\psi^{\varepsilon}_s\rangle
+\langle\psi^{\varepsilon}_s,D_{\mu(s)},\frac{i}{\varepsilon}D_{\mu(s)}\psi^{\varepsilon}_s\rangle\right|
\\&&\leq(\partial_s\mu(s))\|A\|_\infty \|\mathds{1}_{\mathcal{S}_A}\psi^{\varepsilon}_s\|\;.
\end{eqnarray*}
Integrating and observing that $\|D_0\chi\|<\infty$ implies $|\langle\chi,D_0,\chi\rangle|<\infty$, we obtain
$|\langle\psi^{\varepsilon}_s,D_{\mu(s)},\psi^{\varepsilon}_s\rangle|<C $. Using this we get similarly that
$|\langle\psi^{\varepsilon}_s,D^2_s,\psi^{\varepsilon}_s\rangle|<C$, hence with (\ref{ddiff})
\begin{eqnarray}\label{energybound}
\|D_{\mu(s)}A\psi^\varepsilon_u\|&\leq\|(\mu(s)-\mu(u))A^2\psi^\varepsilon_u\|+\|AD_{\mu(u)}\psi^\varepsilon_u\|+\|\sum_{j=1}^3\alpha_j
(\partial_j A)\psi^\varepsilon_u\|
\nonumber\\&\hspace{-0.5cm}\leq(\mu(s)-\mu(u))\|A\|_\infty^2+\|A\|_\infty
\|D_{\mu(u)}\psi^\varepsilon_u\|+\|\nabla A\|_\infty\leq C\;.
\end{eqnarray}
For (\ref{props}) we wish to apply now Corollary \ref{propest} to the first and third term. To apply it to the
first term
$$\|\mathds{1}_\mathcal{S}V_{\mu(s)}^\varepsilon(s,0)\chi\|$$ we need that
$s>\varepsilon(\mu(s)-1)^{-\frac{3}{2(1-\xi)}}\geq \varepsilon(\underline{C}s)^{-\frac{3}{2(1-\xi)}}$, i.e. that
$s^{1+\frac{3}{2(1-\xi)}}>C_\xi\varepsilon$. But since $s>\varepsilon^{1/3-\xi}$ we have that
$s^{1+\frac{3}{2(1-\xi)}}>\varepsilon^{\left(1+\frac{3}{2(1-\xi)}\right)\left(1/3-\xi\right)}$. Since
$\left(1+\frac{3}{2(1-\xi)}\right)\left(1/3-\xi\right)=\frac{5-2\xi}{6}\;\frac{1-3\xi}{1-\xi}\leq\frac{5}{6}<1$
the condition for the Corollary is fulfilled provided that $\varepsilon$ is small enough
($C_\xi\varepsilon^{1/6}<1$). Hence
\begin{equation}\label{firstterm}\|\mathds{1}_\mathcal{S}V_{\mu(s)}^\varepsilon(s,0)\chi\|\leq C_\xi
s^{-1/2}\varepsilon^{\frac{3}{2}-\frac{3}{2}\xi}s^{-\frac{3}{2}+\frac{3}{2}\xi}\;.
\end{equation}
To apply the Corollary to the third term of (\ref{props}) we need that $$s-u>\varepsilon
(\mu(s)-1)^{\frac{3}{2(1-\xi)}}>C_\xi\varepsilon s^{\frac{3}{2(1-\xi)}}\;.$$ Choosing
\begin{equation}\label{sigmatilde}\widetilde{\sigma}=C_\xi\varepsilon s^{\frac{3}{2(1-\xi)}}\end{equation} this is satisfied for all
$u<s-\widetilde{\sigma}$, i.e. for the integrand of the third summand. Hence we have for the third term
\begin{eqnarray*}&&\hspace{-1cm}\frac{1}{\varepsilon}\int_{0}^{s-\widetilde{\sigma}}\left(\mu(s)-\mu(u)\right)\|\mathds{1}_\mathcal{S}V_{\mu(s)}^\varepsilon(s,u)A(x)\psi_u^\varepsilon\|
\\&&\leq\frac{C_\xi}{\varepsilon}\int_{0}^{s-\widetilde{\sigma}}\left(\mu(s)-\mu(v)\right)
s^{-1/2}\varepsilon^{\frac{3}{2}-\frac{3}{2}\xi}(s-v)^{-\frac{3}{2}+\frac{3}{2}\xi} dv
\\&&\leq\frac{C_\xi}{\varepsilon}\int_{0}^{s}
s^{-1/2}\varepsilon^{\frac{3}{2}-\frac{3}{2}\xi}(s-v)^{-\frac{1}{2}+\frac{3}{2}\xi} dv
\\&&\leq C_\xi \varepsilon^{\frac{1}{2}-\frac{3}{2}\xi}s^{\frac{3}{2}\xi}\;.
\end{eqnarray*}
This and (\ref{boundonsecond}) with (\ref{sigmatilde}) introduced and (\ref{firstterm}) in (\ref{props}) yields
\begin{eqnarray*}
\|\mathds{1}_\mathcal{S}\psi^{\varepsilon}_s\|&\leq&C_\xi
\varepsilon^{\frac{3}{2}-\frac{3}{2}\xi}s^{-2+\frac{3}{2}\xi}
+C_\xi\varepsilon s^{-\frac{3}{1-\xi}}
+C_\xi \varepsilon^{\frac{1}{2}-\frac{3}{2}\xi}s^{\frac{3}{2}\xi}
\\&=&C_\xi\varepsilon^{\frac{1}{2}-\frac{3}{2}\xi}s^{-\frac{3}{2}}
\left(\varepsilon s^{-\frac{1}{2}+\frac{3}{2}\xi}
+\varepsilon^{\frac{1}{2}+\frac{3}{2}\xi} s^{\frac{-3-3\xi}{2-2\xi}}
+s^{\frac{3}{2}+\frac{3}{2}\xi}\right)\;.
\end{eqnarray*}
Since $\sigma>s>\varepsilon^{1/3-\xi}$ it follows that for $\varepsilon$ small enough
\begin{eqnarray*}
\varepsilon s^{-\frac{1}{2}+\frac{3}{2}\xi}&\leq&\varepsilon \varepsilon^{-\frac{1}{6}}\sigma^{\frac{3}{2}\xi}<1
\\\varepsilon^{\frac{1}{2}+\frac{3}{2}\xi} s^{\frac{-3-3\xi}{2-2\xi}}&<&\varepsilon^{\frac{1}{2}+\frac{3}{2}\xi}
\varepsilon^{-\frac{1+\xi}{2}\frac{1-3\xi}{1-\xi}}\leq\varepsilon^{\frac{1}{2}+\frac{3}{2}\xi}
\varepsilon^{-\frac{1+\xi}{2}}=\varepsilon^{\xi}<1
\\s^{\frac{3}{2}+\frac{3}{2}\xi}&<&\sigma^{\frac{3}{2}+\frac{3}{2}\xi}<C\;.
\end{eqnarray*}
Hence
\begin{eqnarray*}
\|\mathds{1}_\mathcal{S}\psi^{\varepsilon}_s\|&\leq&C_\xi\varepsilon^{\frac{1}{2}-\frac{3}{2}\xi}s^{-\frac{3}{2}}\;.
\end{eqnarray*}
$\square$

\subsection{Propagation Estimates for the Time Dependent Case: ``Long'' Times}\label{secpe2}

Lemma \ref{timeprop} gives estimates on the decay behavior for times
smaller than $\sigma$. In principle the Lemma can be extended also
for larger times for a very large class of potentials $A_{\mu(s)}$.
This seems alright as long as the propagator
$V_{\mu(s)}^\varepsilon$ leads to fast enough decay, i.e. as long as
$\mu(s)$ is bounded away from one.

But we are especially interested in the case, that $\mu(s)$ attains the critical value $\mu(s)=1$ again after
time $\sigma$, since the potential will be switched off again. We shall need a different technique to estimate
the decay behavior in this situation for times $s>\sigma$ (c.f. Lema \ref{timeprop2}). This techniques will be
based on the fact that by time  $\sigma$ most of the wavefunction has already left the area $\mathcal{S}_A$ of
the potential. This allows us to chose in the comparison of $U^\varepsilon(s,0)\chi$ with
$V_v^{\varepsilon}\chi$ a fixed value of $v$, in fact we shall use $v= \sigma$, in contrast to Lemma
\ref{timeprop} where we chose $v=s$. This has the advantage, that we can use fixed cutoffs in Fourier space,
i.e. we can use Corollary \ref{propest} (i) and (ii).
\begin{lem}\label{timeprop2}(Propagation Estimates - Time Dependent Case: ``Long'' times)\\
 Let $\chi\in L^2$ be normalized
with compact support and finite energy $\|D_0\chi\|<\infty$. Let
$\overline{C}\ge\partial_s \mu(s)\geq \underline{C}>0$ for all
$s\in(0,\sigma)$. Then there exists a constant $C$ such that for
any $0<\xi<1/3$  and all $s\ge\sigma$
\begin{equation}\label{lemmaerste}
\|\mathds{1}_\mathcal{S}U^\varepsilon(s,0)\chi\|\leq
C\varepsilon^{\frac{1}{12}-\frac{3}{4}\xi}\;.\end{equation}
\end{lem}

\noindent {\bf Proof:} Despite the fact that an $L^2$-function has
mostly left any compact region by time $\sigma$, to show that it
scatters is still not easy. The reason is that we deal with a time
evolution which is generated by a time dependent Hamiltonian. We
shall use again a freezing of the potential defining an auxiliary
time evolution. We start with an auxiliary lemma about the auxiliary
time evolution with which we shall later compare the true evolution:
\begin{lem}\label{auxlem}(Auxiliary Lemma)
Let $\widetilde{U}^\varepsilon(s)$ be the unitary defined by $\widetilde{U}^\varepsilon(s,0)=U^\varepsilon(s,0)$
for $s\leq\sigma$ and $\widetilde{U}^\varepsilon(s,\sigma)=V_{\mu(\sigma)}^\varepsilon(s,\sigma)$ for
$s>\sigma$. Let
\begin{equation}\label{chitildeU}
\chi^{\varepsilon}_s:=\widetilde{U}^\varepsilon(s,0)\chi.
\end{equation}
 Then there exists a $\widetilde{\psi}^\varepsilon_s$ such that
\begin{eqnarray}\label{psiminustilde}
\|\chi^{\varepsilon}_s-\widetilde{\psi}^{\varepsilon}_s\|&\leq&C\varepsilon^{\frac{1}{12}-\frac{3}{4}\xi}\;.
\end{eqnarray}and such that for
any $0<\xi<1/3$ and any $\widetilde{m}\in\mathbb{N}$ there exists $C_{\xi,\widetilde{m}}$ such that
\begin{eqnarray}\label{expdecay}
\|\mathds{1}_\mathcal{S} \widetilde{\psi}^{\varepsilon}_s\|&\leq&
C_{\xi,\widetilde{m}}\varepsilon^{\widetilde{m}/3-1}s^{-\widetilde{m}}\;.
\end{eqnarray}

\end{lem}
\noindent {\bf Proof:} With
 the notation (\ref{psiii})
$\chi^{\varepsilon}_s=\psi_s^{\varepsilon}$ for $s\leq\sigma$ and
$\chi^{\varepsilon}_s=V_{\mu(\sigma)}^\varepsilon(s,\sigma)\chi_{\sigma}^{\varepsilon}$
for $s>\sigma$. Using (\ref{cookwave}) with $u=\sigma$ we obtain
$$\chi^{\varepsilon}_\sigma=V_{\mu(\sigma)}^\varepsilon(\sigma,0)\chi
+\frac{i}{\varepsilon}\int_{0}^\sigma\left(\mu(v)-\mu(\sigma)\right)V_{\mu(\sigma)}^\varepsilon(\sigma,v)A(x)\psi_v^{\varepsilon}dv\;.
$$
Hence applying  $V_{\mu(\sigma)}^\varepsilon(s,\sigma)$ yields

\begin{eqnarray}\label{zeitconstant}
\chi^{\varepsilon}_s&=&V_{\mu(\sigma)}^\varepsilon(s,0)\chi
+\frac{i}{\varepsilon}\int_{0}^\sigma\left(\mu(v)-\mu(\sigma)\right)
V_{\mu(\sigma)}^\varepsilon(s,v)A(x)\psi_v^{\varepsilon}dv\;
\nonumber
\\&=&V_{\mu(\sigma)}^\varepsilon(s,0)\chi
+\frac{i}{\varepsilon}\int_{0}^{\sigma-\varepsilon^{2/3}}\left(\mu(v)-\mu(\sigma)\right)
V_{\mu(\sigma)}^\varepsilon(s,v)A(x)\psi_v^{\varepsilon}dv
\nonumber\\&&+\frac{i}{\varepsilon}\int_{\sigma-\varepsilon^{2/3}}^\sigma
\left(\mu(v)-\mu(\sigma)\right)V_{\mu(\sigma)}^\varepsilon(s,v)A(x)\psi_v^{\varepsilon}dv\;.
\end{eqnarray}
The splitting of the integrals are done for application of Corollary \ref{propest} (i) and (ii) to control
(\ref{zeitconstant}) and will become clearer in a moment. We must process in various steps. We define (in view
of Corollary \ref{propest}) now the function $\widetilde{\psi}^{\varepsilon}_s$ of lemma \ref{auxlem}.
\begin{eqnarray}\label{defpsitilde}
\widetilde{\psi}^{\varepsilon}_s&:=&V_{\mu(\sigma)}^\varepsilon(s,0)\rho_{\underline{\kappa},\mu(\sigma)}\chi
\\\nonumber&&+\frac{i}{\varepsilon}\int_{0}^{\sigma-\varepsilon^{2/3}}\left(\mu(v)-\mu(\sigma)\right)V_{\mu(\sigma)}^\varepsilon(s,v)\rho_{\underline{\kappa},\mu(\sigma)}A\psi_v^{\varepsilon}dv
\end{eqnarray}
We note that by definition
\begin{equation}\label{Utildepsitilde}
\widetilde{\psi}^{\varepsilon}_s
=\widetilde{U}^\varepsilon(s,\sigma)\widetilde{\psi}^{\varepsilon}_\sigma\,.
\end{equation}

Now

\begin{eqnarray}\label{defpsitildenorm}
\|\mathds{1}_\mathcal{S}\widetilde{\psi}^{\varepsilon}_s\|&\le&\|\mathds{1}_\mathcal{S}
V_{\mu(\sigma)}^\varepsilon(s,0)\rho_{\underline{\kappa},\mu(\sigma)}\chi\|
\\\nonumber&&+\frac{i}{\varepsilon}\int_{0}^{\sigma-\varepsilon^{2/3}}
\left(\mu(v)-\mu(\sigma)\right)\|\mathds{1}_\mathcal{S}V_{\mu(\sigma)}^\varepsilon(s,v)
\rho_{\underline{\kappa},\mu(\sigma)}A\psi_v^{\varepsilon}\|dv\,.
\end{eqnarray}
We subtract now (\ref{defpsitilde}) from (\ref{zeitconstant}), we take the norms, use triangle inequality and
use unitarity of $V_{\mu(\sigma)}^\varepsilon$
\begin{eqnarray*}
\|\chi^{\varepsilon}_s-\widetilde{\psi}^{\varepsilon}_s\|&\leq&\|\chi-\rho_{\underline{\kappa},\mu(\sigma)}\chi\|
\\\nonumber&&+\frac{1}{\varepsilon}\int_{0}^{\sigma-\varepsilon^{2/3}}|\mu(v)-\mu(\sigma)|\|A\psi_v^{\varepsilon}-\rho_{\underline{\kappa},\mu(\sigma)}A\psi_v^{\varepsilon}\|dv
\nonumber\\&&+C\frac{1}{\varepsilon}\int_{\sigma-\varepsilon^{2/3}}^\sigma|\mu(v)-\mu(\sigma)|\;\|A(x)\psi_v^{\varepsilon}\|dv\;.
\end{eqnarray*}
Using that $\|A(x)\psi_v^{\varepsilon}\|\leq\|A\|_\infty$ one gets after trivial reordering
\begin{eqnarray}\label{massofchi}
\|\chi^{\varepsilon}_s-\widetilde{\psi}^{\varepsilon}_s\|&\leq&\left\|\left(1-\rho_{\underline{\kappa},\mu(\sigma)}\right)\chi\right\|+C\varepsilon^{1/3}
\nonumber\\&&+\frac{C}{\varepsilon}\int_{0}^{\sigma-\varepsilon^{2/3}}(\sigma-v)\left\|\left(1-\rho_{\underline{\kappa},\mu(\sigma)}\right)A\psi_v^{\varepsilon}\right\|dv
\;.
\end{eqnarray}
We shall now estimate the terms in (\ref{defpsitildenorm}) and
(\ref{massofchi}) using Corollary \ref{propest}. The terms are
$\|\mathds{1}_\mathcal{S}V_{\mu(\sigma)}^\varepsilon(s,0)\rho_{\underline{\kappa},\mu(\sigma)}\chi\|$,
$\|(1-\rho_{\underline{\kappa},\mu(\sigma)})\chi\|$,
$\|\mathds{1}_\mathcal{S}V_{\mu(\sigma)}^\varepsilon(s,v)\rho_{\underline{\kappa},\mu(\sigma)}A\psi_v^{\varepsilon}\|$
and
 $\|(1-\rho_{\underline{\kappa},\mu(\sigma)})A\psi_v^{\varepsilon}\|$.

Note that $\chi$  and $A\psi_v^{\varepsilon}$ are compactly
supported and have finite energy (by (\ref{energybound}) and the
assumptions of the lemma). For application of the Corollary
\ref{propest} we must check whether the inequality for the
propagation time (i.e. $s\geq
u>\varepsilon(\mu(v)-1)^{-\frac{3}{2(1-\xi)}}$) is satisfied.

We first want to use the Corollary \ref{propest} (i) and (ii) on
$\chi$ with the following replacements of variables:
$s\widehat{=}s$, $v\widehat{=}\sigma$ and $u\widehat{=}\sigma$.
Hence the condition of the Corollary reads now $s\geq\sigma
>\varepsilon(\mu(\sigma)-1)^{-\frac{3}{2(1-\xi)}}$. The first
inequality is satisfied by assumption of the lemma. Since
$\partial_v \mu(v)\geq \underline{C}>0$ for all $0<v<\sigma$ (by
assumption of the lemma) and $\mu(0)=1$ we have that
$\mu(\sigma)-1>0$. Hence for small enough $\varepsilon$ we have
that $\sigma
>\varepsilon(\mu(\sigma)-1)^{-\frac{3}{2(1-\xi)}}$ and
Corollary \ref{propest} (i) and (ii) yields, observing the
replacements
\begin{eqnarray}\label{tildecor1}
\|\mathds{1}_\mathcal{S}
V_{\mu(\sigma)}(s,0)\rho_{\underline{\kappa},\mu(\sigma)}\chi\|
&\leq&C_{\xi,\widetilde{m}}(\|D_{\mu(\sigma)}\chi\|)\varepsilon^{\widetilde{m}}
s^{-\widetilde{m}}
\end{eqnarray}
and
\begin{eqnarray}\label{tildecor2}
\|(1-\rho_{\underline{\kappa},\mu(\sigma)})\chi\|\leq C
\sigma^{-\frac{3}{4}(1-\xi)}\varepsilon^{\frac{3}{4}(1-\xi)}\sigma^{-1/4}\|\chi\|=
\sigma^{-1+\frac{3}{4}\xi}\varepsilon^{\frac{3}{4}(1-\xi)}\;.
\end{eqnarray}
Next we want to use Corollary \ref{propest} (i) and (ii) replacing $\chi$  by $A(x)\psi_v^{\varepsilon}$ with
$v\leq\sigma-\varepsilon^{2/3}$ where we must make the following replacements of variables in the corollary:
$v\widehat{=}\sigma$, $u\widehat{=}\sigma-v$ and $s\widehat{=}s-v$. Then the condition of the Corollary becomes
$s-v\geq \sigma-v>\varepsilon (\mu(\sigma)-1)^{-\frac{3}{2(1-\xi)}}$,  which is why we did the splitting of the
integrals in (\ref{zeitconstant}) in the first place, namely we have that $v\leq\sigma-\varepsilon^{2/3}$,  so
that the condition is satisfied for small enough $\varepsilon$. Hence we can use the Corollary on
$A(x)\psi_v^{\varepsilon}$ making the correct replacements to obtain
\begin{eqnarray}\label{tildecor3}
\|\mathds{1}_\mathcal{S}
V_{\mu(\sigma)}(s,v)\rho_{\underline{\kappa},\mu(\sigma)}A\psi_v^{\varepsilon}\|
&\leq&C_{\xi,\widetilde{m}}(\|D_{\mu(\sigma)}
A(x)\psi_v^{\varepsilon}\|)(s-v)^{-\widetilde{m}}\varepsilon^{\widetilde{m}}
\end{eqnarray}
and
\begin{eqnarray}\label{tildecor4}
\|(1-\rho_{\underline{\kappa},\mu(\sigma)})A\psi_v^{\varepsilon}\|\leq
C\varepsilon^{\frac{3}{4}(1-\xi)}(\sigma-v)^{-\frac{3}{4}(1-\xi)}\sigma^{-\frac{1}{4}}\|A(x)\psi_v^{\varepsilon}\|\,.
\end{eqnarray}
(\ref{tildecor1})-(\ref{tildecor4}) can now be used to control
(\ref{defpsitildenorm}) and (\ref{massofchi}). Inserting
(\ref{tildecor2}) and (\ref{tildecor4}) into (\ref{massofchi})
yields
\begin{eqnarray*}
\|\chi^{\varepsilon}_s-\widetilde{\psi}^{\varepsilon}_s\|&\leq&C
\sigma^{-1+\frac{3}{4}\xi}\varepsilon^{\frac{3}{4}(1-\xi)}+C\varepsilon^{1/3}
\\&&+\frac{1}{\varepsilon}\int_{0}^{\sigma-\varepsilon^{2/3}}(\sigma-v)(\sigma-v)^{-\frac{3}{4}(1-\xi)}\sigma^{-\frac{1}{4}}\varepsilon^{\frac{3}{4}(1-\xi)}\|A\psi_v^{\varepsilon}\|
dv\,.
\end{eqnarray*}
Now comes Lemma \ref{timeprop} into play. Without the  control of $\|A\psi_v^{\varepsilon}\|$ which the lemma
provides us with, the last summand would be of order $\varepsilon^{-1/4}$ and thus explodes as
$\varepsilon\to0$. But the estimates of Lemma \ref{timeprop} are only good for times larger than
$\varepsilon^{1/3}$. For smaller times the trivial estimate $\|A\psi_v^{\varepsilon}\|\leq C$ is better. Thus we
split the $v$ integral accordingly and arrive at
\begin{eqnarray*}
\|\chi^{\varepsilon}_s-\widetilde{\psi}^{\varepsilon}_s\|&\leq&
\sigma^{-1+\frac{3}{4}\xi}\varepsilon^{\frac{3}{4}(1-\xi)}+C\varepsilon^{1/3}\\
&&+C\frac{1}{\varepsilon}\int_{0}^{\varepsilon^{1/3}}(\sigma-v)^{\frac{1}{4}
+\frac{3}{4}\xi}\varepsilon^{\frac{3}{4}(1-\xi)}\sigma^{-1/4}\|A\psi_v^{\varepsilon}\|
dv
\\&&+C\frac{1}{\varepsilon}\int_{\varepsilon^{1/3}}^{\sigma}(\sigma-v)^{\frac{1}{4}+\frac{3}{4}\xi}\varepsilon^{\frac{3}{4}(1-\xi)}\sigma^{-1/4}\|A\psi_v^{\varepsilon}\|
dv\,.
\end{eqnarray*}
Now use Lemma \ref{timeprop} on the last summand to get (estimating
$\sigma-v\le \sigma\le C$)  that for all $s\ge\sigma$
\begin{eqnarray*}
\|\chi^{\varepsilon}_s-\widetilde{\psi}^{\varepsilon}_s\|&\leq&\sigma^{-1+\frac{3}{4}\xi}\varepsilon^{\frac{3}{4}(1-\xi)}+C\varepsilon^{1/3}
+C\frac{1}{\varepsilon}\sigma^{\frac{3}{4}\xi}\varepsilon^{\frac{3}{4}(1-\xi)}\varepsilon^{1/3}
dv
\nonumber\\&&+C\frac{1}{\varepsilon}\int_{\varepsilon^{1/3}}^{\sigma}(\sigma-v)^{\frac{1}{4}+\frac{3}{4}\xi}\varepsilon^{\frac{3}{4}(1-\xi)}\sigma^{-1/4}
\left(\varepsilon^{\frac{1}{2}-\frac{3}{2}\xi}v^{-\frac{3}{2}}\right)
dv
\nonumber\\&\leq&C\varepsilon^{\frac{3}{4}-\frac{3}{4}\xi}+C\varepsilon^{1/3}
+C\varepsilon^{\frac{1}{12}-\frac{3}{4}\xi}
+C\varepsilon^{\frac{1}{12}-\frac{9}{4}\xi}
\nonumber\\&\leq&C\varepsilon^{\frac{1}{12}-\frac{3}{4}\xi}\;,
\end{eqnarray*}
which is (\ref{psiminustilde}).

Next we estimate (\ref{defpsitildenorm}). Introducing (\ref{tildecor1}) and (\ref{tildecor3}) yields
\begin{eqnarray}\label{einsspsi}
\|\mathds{1}_\mathcal{S} \widetilde{\psi}^{\varepsilon}_s\| &\leq&
C_{\xi,\widetilde{m}}(\|D_{\mu(\sigma)}\chi\|)\varepsilon^{\widetilde{m}}
s^{-\widetilde{m}}
\nonumber\\&&+\frac{C}{\varepsilon}\int_{0}^{\sigma-\varepsilon^{2/3}}\left(\sigma-v\right)
C_{\xi,\widetilde{m}}(\|D_{\mu(\sigma)}A\psi_v^{\varepsilon}\|)\varepsilon^{\widetilde{m}}
\left(s-v\right)^{-\widetilde{m}}dv\;.
\end{eqnarray}
Recall that $s\geq\sigma$, so for $\varepsilon$ small enough we
have
$$s(1-\varepsilon^{2/3}\sigma^{-1})\geq\sigma(1-\varepsilon^{2/3}\sigma^{-1})\;,$$
hence $$s-\sigma+\varepsilon^{2/3}\geq
\varepsilon^{2/3}s\sigma^{-1}\;,$$ so for
$v\leq\sigma-\varepsilon^{2/3}$
$$\left(s-v\right)^{-\widetilde{m}}\leq(s-\sigma+\varepsilon^{2/3})^{-\widetilde{m}}\leq\varepsilon^{-\frac{2}{3}\widetilde{m}}s^{-\widetilde{m}}\sigma^{\widetilde{m}}\;.$$
Using this and the fact that
$\|D_{\mu(\sigma)}A\psi_v^{\varepsilon}\|$ is bounded (c.f.
(\ref{energybound})) we get for (\ref{einsspsi})
\begin{eqnarray*}
\|\mathds{1}_\mathcal{S} \widetilde{\psi}^{\varepsilon}_s\|&\leq&
C_{\xi,\widetilde{m}}\varepsilon^{\widetilde{m}} s^{-\widetilde{m}}
+C_{\xi,\widetilde{m}}\varepsilon^{\widetilde{m}/3-1}s^{-\widetilde{m}}\leq
2C_{\xi,\widetilde{m}}\varepsilon^{\widetilde{m}/3-1}s^{-\widetilde{m}}\;,
\end{eqnarray*}
which is (\ref{expdecay}).

 $\square$

We shall now prove Lemma \ref{timeprop2}. Using ( \ref{cook}) we have for $s>\sigma$ that
$$
\left(U^\varepsilon(s,\sigma)-\widetilde{U}(s,\sigma)\right)\widetilde{\psi}^{\varepsilon}_\sigma
=-\frac{i}{\varepsilon}\int_{\sigma}^sU^\varepsilon(s,v)\left(\mu(\sigma)-\mu(v)\right)A(x)\widetilde{U}^\varepsilon(v,\sigma)\widetilde{\psi}^{\varepsilon}_\sigma
dv
$$
and therefore  by (\ref{Utildepsitilde})
\begin{eqnarray*}
\|\mathds{1}_\mathcal{S}U^\varepsilon(s,\sigma)\widetilde{\psi}^{\varepsilon}_\sigma\|
&\leq&\|\mathds{1}_\mathcal{S}\widetilde{\psi}^{\varepsilon}_s\|
+\frac{1}{\varepsilon}\int_{\sigma}^s\left(\mu(v)-\mu(\sigma)\right)\|A
\|_\infty\|\mathds{1}_\mathcal{S}\widetilde{\psi}^{\varepsilon}_v\|
dv\,.
\end{eqnarray*}
Using (\ref{expdecay})
\begin{eqnarray}\label{expdecay2}
\|\mathds{1}_\mathcal{S}U^\varepsilon(s,\sigma)\widetilde{\psi}^{\varepsilon}_\sigma\|&\leq&
C\varepsilon^{\widetilde{m}}s^{-\widetilde{m}}+\frac{C}{\varepsilon}\int_{\sigma}^s\left(\mu(\sigma)-\mu(v)\right)v^{-\widetilde{m}}\varepsilon^{\widetilde{m}}
dv
\nonumber\\&\leq&C\varepsilon^{\widetilde{m}}s^{-\widetilde{m}}+\frac{C}{\varepsilon}\int_{\sigma}^sv^{-\widetilde{m}+1}\varepsilon^{\widetilde{m}}
dv
\nonumber\\&\leq&C\varepsilon^{\widetilde{m}}s^{-\widetilde{m}}+C(s^{-\widetilde{m}+2}-\sigma^{-\widetilde{m}+2})\varepsilon^{\widetilde{m}-1}
\nonumber\\&\leq&
C\varepsilon^{\widetilde{m}}s^{-\widetilde{m}}+Cs^{-\widetilde{m}+2}\varepsilon^{\widetilde{m}-1}\leq
Cs^{-\widetilde{m}}\varepsilon^{\widetilde{m}-1}\;.
\end{eqnarray}
We turn now to
$\left\|\mathds{1}_\mathcal{S}U^\varepsilon(s,0)\chi\right\|$.
Recall that
$\chi^{\varepsilon}_\sigma=\widetilde{U}^\varepsilon(\sigma,0)\chi=U^\varepsilon(\sigma,0)\chi$,
thus
\begin{eqnarray*}\left\|\mathds{1}_\mathcal{S}U^\varepsilon(s,0)\chi\right\|
&=&\left\|\mathds{1}_\mathcal{S}U^\varepsilon(s,\sigma)\chi^{\varepsilon}_\sigma\right\|
\\&\leq&
\left\|\mathds{1}_\mathcal{S}U^\varepsilon(s,\sigma)\left(\widetilde{\psi}^{\varepsilon}_\sigma-
\chi^{\varepsilon}_\sigma\right)\right\|
+\left\|\mathds{1}_\mathcal{S}U^\varepsilon(s,\sigma)\widetilde{\psi}^{\varepsilon}_\sigma\right\|\\
&\leq&
\left\|\widetilde{\psi}^{\varepsilon}_\sigma-\chi^{\varepsilon}_\sigma\right\|
+\left\|\mathds{1}_\mathcal{S}U^\varepsilon(s,\sigma)\widetilde{\psi}^{\varepsilon}_\sigma\right\|\\
&\leq&C\varepsilon^{\frac{1}{12}-\frac{3}{4}\xi}+
Cs^{-\widetilde{m}}\varepsilon^{\widetilde{m}-1}\;,
\end{eqnarray*}
where we used (\ref{psiminustilde}) and (\ref{expdecay2}).
Choosing $\widetilde{m}=2$ the Lemma follows.
\subsection{Control of $\psi_{s}^\varepsilon$ for $ s>
0$}\label{endofproof}

We come now to the proof of Theorem \ref{onecrit}. We wish to
establish that for $s>0$ and $\chi \in L^2:$ $\lim_{\varepsilon\to
0}\langle \psi_s^{\varepsilon},\chi\rangle=0.$ From Lemma
\ref{corvor0} we have that
$\lim_{\varepsilon\to0}\|(1-P_\mathcal{N}
)\psi_0^{\varepsilon}\|=0.$ Therefore by
$$\lim_{\varepsilon\to
0}\langle \psi_s^{\varepsilon},\chi\rangle=\lim_{\varepsilon\to
0}\langle U^\varepsilon(s,0)P_\mathcal{N}
\psi_0^{\varepsilon},\chi\rangle + \lim_{\varepsilon\to 0}\langle
U^\varepsilon(s,0)(1-P_\mathcal{N}
)\psi_0^{\varepsilon},\chi\rangle,$$ and
$$\lim_{\varepsilon\to 0}\langle
U^\varepsilon(s,0)(1-P_\mathcal{N}
)\psi_0^{\varepsilon},\chi\rangle\le\lim_{\varepsilon\to
0}\|(1-P_\mathcal{N} )\psi_0^{\varepsilon}\|=0 $$ Theorem
\ref{onecrit} follows from
\begin{cor}\label{lemmagrosss}(Decay of the Critical Bound State)\\
Let $s>0$ and $\chi\in L^2$. Then
\begin{equation}\label{onecriteqb2}
\lim_{\varepsilon\to0}|\langle U^\varepsilon(s,0)P_\mathcal{N}
\psi_0^{\varepsilon},\chi\rangle|=0\;.
\end{equation}
\end{cor}
 \noindent\textbf{Proof:}
Note that $P_\mathcal{N} $ projects on the subspace with energy 1,
hence $\|D_0P_\mathcal{N} \psi_0^{\varepsilon}\|\leq1$.  For the
proof it is very convenient to use a two scale argument. Let
$j(\mathbf{x})\in C^\infty$ be a mollifier with $j(\mathbf{x})=1$
for $x\leq 1$ and $j(\mathbf{x})=0$ for $x\geq 2$, define for any
$\delta>0$ $j_{\delta}(\mathbf{x}):=j(\delta\mathbf{x})$,
$\chi^1_{\delta,\varepsilon}:=j_{\delta}P_\mathcal{N}
\psi_0^{\varepsilon}$ and $\chi^2_{\delta}:=j_{\delta}\chi$. Note
that this definition yields, that
\begin{equation}\label{bothlimits}\lim_{\delta\to0}\|P_\mathcal{N} \psi_0^{\varepsilon}-\chi^1_{\delta,\varepsilon}\|=0=\lim_{\delta\to0}\|\chi-\chi^2_{\delta}\|
\end{equation}
and
\begin{eqnarray*}\|D_0\chi^1_{\delta,\varepsilon}\|&=&\|D_0j_{\delta}P_\mathcal{N} \psi_0^{\varepsilon}\|\infty
\leq C\sup_{k=1,2,3}\|\partial_k
j_{\delta}\|_\infty\|P_\mathcal{N}
\psi_0^{\varepsilon}\|+\|j_{\delta}D_0P_\mathcal{N}
\psi_0^{\varepsilon}\|
\\&=&C\delta+\|j_{\delta}P_\mathcal{N} \psi_0^{\varepsilon}\|\leq
C\delta+\|\psi_0^{\varepsilon}\|<\infty\;.
\end{eqnarray*}
Now let $s>0$. For any $\delta>0$ we can use  Lemma \ref{timeprop}
and Lemma \ref{timeprop2} setting $\xi=1/12$ to get that
$$\|\mathds{1}_\mathcal{S_{\delta}}U^\varepsilon(s,0)\chi^1_{\delta,\varepsilon}\|\leq
C_{\delta} \varepsilon^{\frac{1}{48}}\;,$$where
$\mathcal{S}_\delta$ is the support of $j_\delta$. Hence
\begin{eqnarray*}
|\langle
 U^\varepsilon(s,0)P_\mathcal{N} \psi_0^{\varepsilon},\chi\rangle|
&\leq&|\langle U^\varepsilon(s,0)P_\mathcal{N}
\psi_0^{\varepsilon},\chi^2_{\delta}\rangle|+|U^\varepsilon(s,0)P_\mathcal{N}
\psi_0^{\varepsilon},\chi-\chi^2_{\delta}\rangle|
\\&\leq&|\langle
j_{\delta}U^\varepsilon(s,0)P_\mathcal{N}
\psi_0^{\varepsilon},\chi\rangle|+\|P_\mathcal{N}
\psi_0^{\varepsilon}\|\;\|\chi-\chi^2_{\delta}\|
\\&\leq&\langle
\mathds{1}_\mathcal{S_{\delta}}U^\varepsilon(s,0)P_\mathcal{N}
\psi_0^{\varepsilon},\chi\rangle|+\|\chi-\chi^2_{\delta}\|
\\&\leq&\|\mathds{1}_\mathcal{S_{\delta}}U^\varepsilon(s,0)P_\mathcal{N} \psi_0^{\varepsilon}\|\;
\|\chi\|+\|\chi-\chi^2_{\delta}\|
\\&\leq&\|\mathds{1}_\mathcal{S_{\delta}}U^\varepsilon(s,0)\chi^1_{\delta,\varepsilon}\|
+\|\mathds{1}_\mathcal{S_{\delta}}U^\varepsilon(s,0)\left(P_\mathcal{N}
\psi_0^{\varepsilon}- \chi^1_{\delta,\varepsilon}\right)\|
\\&&+\|\chi-\chi^2_{\delta}\|
\\&=&C_{\delta}
\varepsilon^{\frac{1}{48}}+\|P_\mathcal{N} \psi_0^{\varepsilon}-
\chi^1_{\delta,\varepsilon}\|+\|\chi-\chi^2_{\delta}\|\;.
\end{eqnarray*}
Taking first the limit $\varepsilon\to0$ and then $\delta\to0$ the
Corollary follows in view of (\ref{bothlimits}). $\square$

{\bf Acknowledgement}: We thank  ESI (Vienna) for hospitality and
funds. Work was partly funded by DFG.

\section{Appendix: Proof of Lemma \ref{propzeta} (\ref{l1abschb2})}


Recall (\ref{produkte})
\begin{eqnarray*}
&&\hspace{-0.5cm}V_\mu(t,0)\rho_{\underline{\kappa},\mu}(1-\rho_{\overline{\kappa},\mu})\chi(\mathbf{x})
\\&&=\sum_{j=1}^{4}\int(2\pi)^{-\frac{3}{2}}
 \exp\left(-itE_k\right) \varphi_{\mu}(\mathbf{k},j,\mathbf{x})
 \widehat{\rho}_{\underline{\kappa}}(1-\widehat{\rho}_{\overline{\kappa}})\mathcal{F}_\mu(\chi)(\mathbf{k},j)d^{3}k\;.
\end{eqnarray*}

We estimate the right hand side via stationary phase method, i.e. we
integrate by parts. Using $\frac{i E_k}{kt}\partial_k\exp\left(-i
tE_k\right)=\exp\left(-i tE_k\right)$ m partial integrations yield
 - writing
$$\left(\partial_k\frac{E_k}{k}\right)^m:=\partial_k\frac{E_k}{k} \partial_k\frac{E_k}{k} \ldots\;,$$
where $\partial_k$ acts on everything to the right -
\begin{eqnarray*}
&&\hspace{-1cm}V_\mu(t,0)\rho_{\underline{\kappa},\mu}(1-\rho_{\overline{\kappa},\mu})\chi(\mathbf{x})
\\&&\;=\left(-\frac{i}{t}\right)^m\sum_{j=1}^{4}\int_{0}^\infty\int(2\pi)^{-\frac{3}{2}}
\exp\left(-itE_k\right)
\\&&\;\;\;\left(\left(\partial_k\frac{E_k}{k}\right)^m\varphi_{\mu}(\mathbf{k},j,\mathbf{x})
 \widehat{\rho}_{\underline{\kappa}}(1-\widehat{\rho}_{\overline{\kappa}})\mathcal{F}_\mu(\chi)(\mathbf{k},j)\right)d\Omega dk
\\&&\;=\left(-\frac{i}{t}\right)^m\sum_{j=1}^{4}\int k^{-2}(2\pi)^{-\frac{3}{2}}
\exp\left(-i tE_k\right)
\\&&\;\;\;\left(\left(\partial_k\frac{E_k}{k}\right)^m\varphi_{\mu}(\mathbf{k},j,\mathbf{x})
 \widehat{\rho}_{\underline{\kappa}}(1-\widehat{\rho}_{\overline{\kappa}})\mathcal{F}_\mu(\chi)(\mathbf{k},j)k^2\right) d^3k
\end{eqnarray*}

Since $\widehat{\rho}_{\underline{\kappa}}(\mathbf{k})=0$ for $k\leq
\underline{\kappa}$ and $k\geq K$

\begin{eqnarray}\label{ourcase}
&&\hspace{-0.5cm}\|\mathds{1}_\mathcal{S}
V_\mu(u,0)\chi\|_\infty\leq t^{-m} \frac{4}{3}\pi
\overline{\kappa}^34
\\&& \sup_{2\overline{\kappa}>k>\underline{\kappa},\mathbf{x}\in\mathcal{S},j}\left|\nonumber
k^{-2}\left(\left(
\partial_k\frac{E_k}{k}\right)^m\varphi_{\mu}(\mathbf{k},j,\mathbf{x})
 \widehat{\rho}_{\underline{\kappa}}(1-\widehat{\rho}_{\overline{\kappa}})\mathcal{F}_\mu(\chi)(\mathbf{k},j)k^2\right)\right|_\infty\;.
\end{eqnarray}

We next show that for any $j,l,r\in\mathbb{N}_0$ there exist $C_{j,l,r}$ so that

\begin{eqnarray}\label{va1a}
\left(\partial_k\frac{E_k}{k}\right)^nk^2
f(k)&=&\sum_{j+l+r=m}C_{j,l,r}E_k^{m-2r} k^{-m-l+r+2}
\partial_k^j f(k)\;.
\end{eqnarray}

We prove this equation by induction over $m$. For $m=0$ (\ref{va1a})
follows trivially. Assume that (\ref{va1a}) holds for some
$m\in\mathbb{N}$. It follows that

\begin{eqnarray*}
\left(\partial_k\frac{E_k}{k}\right)^{m+1}k^2 f(k)&=&
\partial_k\frac{E_k}{k}\left(\partial_k\frac{E_k}{k}\right)^nk^2 f(k)
\\&=&\partial_k\frac{E_k}{k}\sum_{j+l+r=m}C_{j,l,r}E_k^{m-2r} k^{-m+2-l+r}\partial_k^j f(k)
\\&=&\partial_k\sum_{j+l+r=m}C_{j,l,r}E_k^{m-2r+1} k^{(-m-1+2)-l+r}\partial_k^j f(k)
\\&=&\sum_{j+l+r=m}C_{j,l,r}\left(\partial_kE_k^{m-2r+1}\right) k^{(-m-1+2)-l+r}\partial_k^j f(k)
\\&&+\sum_{j+l+r=m}C_{j,l,r}E_k^{m-2r+1} \left(\partial_kk^{(-m-1+2)-l+r}\right)\partial_k^j f(k)
\\&&+\sum_{j+l+r=m}C_{j,l,r}E_k^{m-2r+1} k^{-m+3-l+r}\partial_k^{j+1} f(k)
\end{eqnarray*}

Using that $E_k=\sqrt{k^2+1}$ we have that

$$\partial_k E_k ^m = m E_k^{m-1} \partial_k\sqrt{k^2+1}=m E_k^{m-2}k\;.$$

Setting $\widetilde{m}=m+1$, $\widetilde{j}=j+1$,
$\widetilde{l}=l+1$ and $\widetilde{r}=r+1$ yields

\begin{eqnarray*}
\left(\partial_k\frac{E_k}{k}\right)^{m+1}k^2 f(k)
&=&\sum_{j+l+\widetilde{r}=\widetilde{m}}C_{j,l,r}E_k^{\widetilde{m}-2\widetilde{r}}
k^{-\widetilde{m}+2-l+\widetilde{r}}\partial_k^j f(k)
\\&&+\sum_{j+l+r=\widetilde{m}}C_{j,l,r}E_k^{\widetilde{m}-2r} k^{-\widetilde{m}+2-\widetilde{l}+r}\partial_k^j
f(k)
\\&&+\sum_{\widetilde{j}+l+r=\widetilde{m}}C_{\widetilde{j},l,r}E_k^{\widetilde{m}-2r}
k^{-\widetilde{m}+2-l+r}\partial_k^{j+1} f(k)
\end{eqnarray*}

for appropriate $C_{\widetilde{j},l,r}<\infty$,
$C_{j,\widetilde{l},r}<\infty$ and $C_{j,l,\widetilde{r}}<\infty$,
and (\ref{va1a}) follows  for $\widetilde{m}=m+1$. Induction yields
that (\ref{va1a}) holds for all $m\in\mathbb{N}_0$.

Note that for $k\to 0 $\\$k^{-2}E_k^{m-2r} k^{-m+2-l+r}$ is of order
$k^{-m-l+r}$. For $k\to\infty$ $E_k$ is of order $k$, hence $k^{-2}
E_k^{m-2r} k^{-m+2-l+r}$ is of order $k^{-l-r}$ (hence bounded for
large $k$). Since we only observe $\underline{\kappa}\to0$ it
follows with (\ref{va1a}) that for any $m,j\in\mathbb{N}_0$ there
exist $C_{m,j}<\infty$ such that

\begin{eqnarray}\label{va1}
\mid k^2\left(\partial_k\frac{E_k}{k}\right)^nk^2 f(k)\mid&
\leq&\sum_{j=0}^m C_{m,j} k^{-2m+j} \mid \partial_k^j f(k)\mid\;.
\end{eqnarray}
In our case (c.f. (\ref{ourcase})) we have $f=\varphi_{\mu}
 \widehat{\rho}_{\underline{\kappa}}(1-\widehat{\rho}_{\overline{\kappa}})\mathcal{F}_\mu(\chi)$.
Using the product rule of differentiation it follows that
\begin{eqnarray*}
&&\hspace{-0.5cm}\partial_k^j\varphi_{\mu}
 \widehat{\rho}_{\underline{\kappa}}(1-\widehat{\rho}_{\overline{\kappa}})\mathcal{F}_\mu(\chi)
\\&&=\sum_{j_1+j_2+j_3+j_4=j}C_{j_1,j_2,j_3,j_4}\left(\partial_k^{j_1}\varphi_{\mu}\right)
\left(\partial_k^{j_2}
\widehat{\rho}_{\underline{\kappa}}\right)\left(1-\partial_k^{j_3}\widehat{\rho}_{\overline{\kappa}}\right)\left(\partial_k^{j_4}\mathcal{F}_\mu(\chi)\right)
 \;,
\end{eqnarray*}
where $C_{j_1,j_2,j_3,j_4}$ is a combinatorial factor. With (\ref{defrho2}), (\ref{31}) and (\ref{32}) we get
using that $\underline{\kappa}<\overline{\kappa}$
\begin{eqnarray*}
|\partial_k^j\varphi_{\mu}
 \widehat{\rho}_{\underline{\kappa}}(1-\widehat{\rho}_{\overline{\kappa}})\mathcal{F}_\mu(\chi)|
&<&\sum_{j_1+j_2+j_3+j_4=j}C_{j_1,j_2,j_3,j_4}C_{j_1}C_{j_4}\underline{\kappa}^{-j_2-j_3}(1+x)^{j_1}
\\&&\hspace{0.5cm}\left(\underline{\kappa}^{-1}+\left|\sum_{l=1}^n\frac{k}{
 |\mu-1- \nu_l k^2|
+ck^3}\right|\right)^{j_1+j_4+2}
 \;.
\end{eqnarray*}
Collecting the worst terms (i.e. handling the two cases $\underline{\kappa}^{-1}<\left|\sum_{l=1}^n\frac{k}{
 |\mu-1- \nu_l k^2|
+ck^3}\right|$ and `` $ \geq $ '' separately) we get with an appropriate constant $C_j$ that
\begin{eqnarray*}&&\hspace{-0.5cm}|\partial_k^j\varphi_{\mu}
 \widehat{\rho}_{\underline{\kappa}}(1-\widehat{\rho}_{\overline{\kappa}})\mathcal{F}_\mu(\chi)|
\\&&<C_j(1+x)^j\left(\underline{\kappa}^{-j-2}+\left|\sum_{l=1}^n\frac{k}{
 |\mu-1- \nu_l k^2|
+ck^3}\right|^{j+2}\right)\;.\end{eqnarray*} With (\ref{va1}), again collecting the worst terms, it follows that
\begin{eqnarray*}
\mid k^2\left(\partial_k\frac{E_k}{k}\right)^nk^2 \varphi_{\mu}
 \widehat{\rho}_{\underline{\kappa}}(1-\widehat{\rho}_{\overline{\kappa}})\mathcal{F}_\mu(\chi)\mid
<(1+x)^m C_m\left(\underline{\kappa}^{-m-2}k^{-m}+k^{-2m}\right)
\\\hspace{2.5cm}+(1+x)^m
C_mk^2\left|\sum_{l=1}^n\frac{1}{
 |\mu-1- \nu_l k^2|
+ck^3}\right|^{m+2}\;.
\end{eqnarray*}
Thus (recall that $\underline{\kappa}<1$ and that $\mathcal{S}$ is
compactly supported, hence $(1+x)^m$ is bounded by some constant)
\begin{eqnarray}\label{thelatter}
&&\hspace{-1cm}\sup_{2\overline{\kappa}>k>\underline{\kappa},\mathbf{x}\in\mathcal{S},j}\left|
k^{-2}\left(\left(
\partial_k\frac{E_k}{k}\right)^m\varphi_{\mu}(\mathbf{k},j,\mathbf{x})
 \widehat{\rho}_{\underline{\kappa}}(1-\widehat{\rho}_{\overline{\kappa}})
 \mathcal{F}_\mu(\chi)(\mathbf{k},j)k^2\right)\right|_\infty
\\\nonumber&&<C_m\left(\underline{\kappa}^{-2m-2}+\sup_{2\overline{\kappa}>k>\underline{\kappa}}\left(k^2\left|\sum_{l=1}^n\frac{1}{
 |\mu-1- \nu_l k^2|
+ck^3}\right|^{m+2}\right)\right)\;.
\end{eqnarray}
Since $\underline{\kappa}<1$ and thus
$$\sup_{2\overline{\kappa}>k>\underline{\kappa}}\left\{k^2\left|\sum_{l=1}^n\frac{1}{
 |\mu-1- \nu_l k^2|
+ck^3}\right|^2\right\}<\frac{1}{c^2\underline{\kappa}^4}\;,$$
(\ref{thelatter}) is bounded from above by
$$C_m\underline{\kappa}^{-4}\left(\underline{\kappa}^{-2m}+\sup_{2\overline{\kappa}>k>\underline{\kappa}}\left|\sum_{l=1}^n\frac{1}{
 |\mu-1- \nu_l k^2|
+ck^3}\right|^{m}\right)\;.$$ With (\ref{ourcase}) (and using that
for positive $a,b$ and $m\in\mathds{N}$ we have $(a+b)^m\geq
a^m+b^m$) equation (\ref{l1abschb2}) follows.
\begin{flushright}$\Box$\end{flushright}


\begin{thebibliography}{}






\bibitem{bhabba} Bhabba H. J. : The Creation of Electron Pairs by Fast Charged Particles, Proc. R. Soc. London
Ser. A {\bf 152}, 559-586 (1935).



\bibitem{beck} Beck F. , Steinwedel H. and S\"ussmann G.: Bemerkungen zum Klein'schen Paradoxon, Z.Phys {\bf 171}, 189-198
(1963).

\bibitem{itzyk} Brezin  E. and Itzykson C.:
Pair Production in Vacuum by an Alternating Field, Phys. Rev. D.
{\bf 2}, 1191-1199 (1970).


\bibitem{exp2} Cowan T. , et al.:
Observation of correlated narrow-peak structures in positron and
electron spectra from superheavy collision systems, Phys. Rev. Lett.
{\bf 56}, 444--447 (1986).

\bibitem{diracbook} Dirac P.: The Principles of Quantum Mechanics,
Oxford University Press, Oxford (1930).

\bibitem{loss} Dolbeault J., Esteban M. J.  and Loss M.:
Relativistic hydrogenic atoms in strong magnetic fields,
arXiv:math/0607027v1 [math.AP].

\bibitem{duerr} D\"urr D.  and Pickl P.:
Flux-across-surfaces Theorem for a Dirac-particle, J. Math. Phys.
{\bf 44}, 423--465 (2003).


\bibitem{gersh} Gershtein S. and Zeldovich Y.: Positron Production During the Mutual Approach of Heavy Nuclei and the Polarization of the Vacuum, Sov. Phys. JETP {\bf 30}, 358-361
(1970).



\bibitem{hainzl} Hainzl C., Lewin M. and Solovej J. P.:
Mean-field approximation in Quantum Electrodynamics. The no-photon
case, to appear in Comm. Pure Appl. Math.



\bibitem{heis} Heisenberg W., Euler H.: Consequences of Dirac's Theory of the Positron, Z. Phys. {\bf 98}, 714
(1936).

\bibitem{klein} Klein O.: Die Reflexion von Elektronen an einem Potentialsprung nach der relativistischen Dynamik von Dirac, Z. Phys. {\bf 53}, 157
(1929).


\bibitem{greiner} Greiner W., M\"uller B. and Rafelski J.:
Quantum Electrodynamics of Strong Fields, Springer Verlag, Berlin
(1985).

\bibitem{ikebe} Ikebe T.: Eigenfunction expansions assoziated with
the Schr\"odinger operators and their application to scattering
theory, Arch. Rat. Mech. Anal. {\bf 5}, 1-34 (1960).

\bibitem{jensen} Jensen A., Kato T.: Spectral Properties of Schr\"odinger operators and time-decay of the wavefunctions, Duke Math. J. {\bf 46} no. 3, 583-611 (1979).


\bibitem{klaus} Klaus M.: On couplin constant thresholds and related eigenvalue properties of Dirac operators,
J. Reine Angew.Math. {\bf 362} 197-212 (1985).


\bibitem{mueller} M\"uller B.:
Positron creation in superheavy quasimolecules, Ann. Rev. Nucl.
Science {\bf 26}, 351--383 (1976).


\bibitem{mprg} M\"uller B., Peitz H., Rafelski J. and Greiner W.:
Solutions of the Dirac Equation for Strong External Fields, Phys.
Rev. Lett. {\bf 28}, 1235--1238 (1972).

\bibitem{PRLmueller} M\"uller B. and Rafelski J.:
Stabilization of the Charged Vacuum Created by Very Strong
Electrical Fields in Nuclear Matter, Phys. Rev. Lett. {\bf 34},
349--352 (1975).



\bibitem{nenciu1} Nenciu G.:
On the adiabatic limit for Dirac particles in external fields,
Commun. Math. Phys. {\bf 76}, 117-128 (1980).

\bibitem{nenciu2} Nenciu G.:
Existence of spontaneous pair creation in the external field
approximation of Q.E.D., Commun. Math. Phys. {\bf 109}, 303-312
(1987).


\bibitem{neutronen} O'Connell R.F.:
Effect of the Anomalous Magnetic Moment of the Electron on
Spontaneous Pair Production in a Strong Magnetic Field, Phys. Rev.
Lett. {\bf 21}, 397--398 (1968).



\bibitem{diss} Pickl P.: Existence of Spontaneous Pair Creation,
Dissertation (2005).


\bibitem{picklneu} Pickl P.:
Generalized Eigenfunctions for Dirac Operators Near Criticality,
arXiv:math-ph/0608004.


\bibitem{PDPhysrev} Pickl P. and D\"urr D.:
 Adiabatic Pair Creation in Heavy Ion and Laser Fields,
 arXiv:hep-th/0609200.





\bibitem{prodan} Prodan E.: Spontaneous transitions in quantum mechanics, J. Phys. A: Math. Gen., {\bf 32} 4877-4881 (1999).

\bibitem{PRLgreiner2} Rafelski J., Fulcher L.P. and Greiner W.:
Superheavy Elements and an Upper Limit to the Electric Field
Strength, Phys. Rev. Lett. {\bf 27}, 958--961 (1971).


\bibitem{reedsimon} Reed M.  und Simon B.:  Functional Analysis, Academic Press, San Diego,
(1980).

\bibitem{rein} Rein D.: \"Uber den Grundzustand \"uberschwerer Atome, Z. Phys {\bf 221}, 423-430
(1969).


\bibitem{reinhardt} Reinhardt  J., U. M\"uller, B. M\"uller and Greiner W: The decay of the vacuum in the field of superheavy nuclear systems, Z. f. Physik A {\bf 303}, 173--188
(1981).




\bibitem{riesz} Riesz F., von Sz.-Nagy B.: Functional Analysis.
New York: F. Ungar. Publ. Co. (1955).



\bibitem{laser} Roberts C.D., Schmidt S.M. and  Vinnik D.V.:
Quantum Effects with an X-Ray Free-Electron Laser, Phys. Rev. Lett.
{\bf 89}, 153901 (2002).

\bibitem{rodni} Rodnianski, I. and Schlag, W.: Time decay for
solutions of Schr\"odinger equations with rough and time dependent
potentials, Invent. math. {\bf 155}, 451-513 (2004).

\bibitem{sauter} Sauter F.: \"Uber das Verhalten eines Elektrons im homogenen elektrischen Feld nach der relativistischen Theorie
Diracs, Z. Phys. {\bf 69}, 742 (1931).

\bibitem{schwinger} Schwinger J.:
On Gauge Invariance and Vacuum Polarization, Phys. Rev. {\bf 82},
664-679 (1951).


\bibitem{scharf} Scharf G., Seipp H.P.: Charged Vacuum, Spontaneous Positron Production and all that, Phys.
Lett. {\bf 108} B, 196--198 (1982).


\bibitem{exp1} Schweppe J., et al.:
Observation of a Peak Structure in Positron Spectra from U+Cm
Collisions, Phys. Rev. Letters {\bf 51}, 2261--2264 (1983).

\bibitem{PRLgreiner} Smith K., Peitz H., M\"uller B. and Greiner W.:
Induced Decay of the Neutral Vaccum in Overcritical Fields Occurring
in Heavy-Ion Collisions, Phys. Rev. Lett. {\bf 32}, 554--556 (1974).


\bibitem{stefan} Teufel S.:
Adiabatic Perturbation Theory in Quantum Dynamics, Springer Verlag,
Berlin (2000).

\bibitem{teufel} Teufel S.:
The flux-across-surfaces theorem and its implications for scattering
theory, Dissertiation an der Ludwig-Maximilians-Universit\"at,
M\"unchen (1999).

\bibitem{thaller} Thaller B.: The Dirac equation, Springer Verlag,
Berlin (1992).






\bibitem{yamada}  Yamada O.:  Eigenfunction expansions and scattering theory for Dirac operators, Publ. RIMS. Kyoto Univ., {\bf 11},
651-689 (1976).








\end{thebibliography}
\end{document}